\documentclass[aps,pra,10pt,a4paper,twocolumn,footinbib, superscriptaddress]{revtex4-2}

\usepackage{graphicx}
\usepackage{amsmath}
\usepackage{amsfonts}
\usepackage{amssymb}
\usepackage{soul}
\usepackage{dsfont}
\usepackage{hyperref}
\usepackage{amstext}
\usepackage[caption=false]{subfig}
\usepackage{epstopdf} 
\usepackage{mathtools} 
\usepackage{balance}
\hypersetup{colorlinks=true, citecolor=blue, urlcolor=blue, linkcolor=blue}
\usepackage{color,soul}
\usepackage{physics}
\usepackage{braket}

\date{\today}

\begin{document}

\title{Enhanced optomechanical nonlinearity through non-Markovian mechanical noise}

\author{Sofia Qvarfort}
\affiliation{Nordita, KTH Royal Institute of Technology and Stockholm University, Hannes Alfv\'{e}ns v\"{a}g 12, SE-106 91 Stockholm, Sweden}
\affiliation{Department of Physics, Stockholm University, AlbaNova University Center, SE-106 91 Stockholm, Sweden}
\email{sofia.qvarfort@fysik.su.se}

\begin{abstract}
Cavity optomechanical systems in the quantum regime consist of a cavity mode and mechanical element coupled together through radiation pressure. In the nonlinear optomechanical regime, open-system dynamics is generally challenging to treat analytically, since the noise terms do not commute with the optomechanical interaction term. Specifically, a general treatment of both Markovian and non-Markovian mechanical noise in the nonlinear optomechanical regime is still outstanding. Here we address this question by solving the full dynamics of an optomechanical system in the nonlinear regime where the mechanical element interacts with a bath of harmonic oscillators, representing full quantum Brownian motion. The solutions, which are exact and analytic, allow us to consider the strength of the optomechanical nonlinearity in the presence of both Markovian (Ohmic) and non-Markovian (sub-Ohmic and super-Ohmic) baths. We show that that while the strength of the nonlinearity is generally reduced by a Markovian bath spectrum, it can be enhanced by constructing a bath with a highly non-Markovian structure. The results have potential implications for future optomechanical experiments which seek to achieve a strong optomechanical nonlinearity.  
\end{abstract}

\maketitle

\section{Introduction} 
The control of mechanical resonators in the quantum regime has seen significant improvements over the last decade, both in terms of theoretical and experimental advances~\cite{aspelmeyer2014cavity}. In particular, the achievements of ground-state cooling for both clamped~\cite{teufel2011sideband, chan2011laser} and levitated systems~\cite{delic2020cooling, piotrowski2023simultaneous} have set the stage for preparations of highly non-classical states beyond the ground state~\cite{bild2023schrodinger}. The relatively large mass of these systems compared with the single-atom scale allows for a number of applications, including quantum-enhanced sensing~\cite{qvarfort2018gravimetry, schneiter2020optimal, qvarfort2021optimal} (see~\cite{rademacher2020quantum} for a review), and  tests of fundamental physics, including gravitational decoherence~\cite{bassi2017gravitational} or gravity-mediated entanglement~\cite{bose2017spin,marletto2017gravitationally}. 

In cavity optomechanical systems, the coupling between the cavity mode and mechanical mode is mediated through radiation pressure. This interaction can be represented with a cubic Hamiltonian term, which gives rise to nonlinear equations of motions. The dynamics of cavity opomechanical systems in the nonlinear regime was first solved in~\cite{bose1997preparation, mancini1997ponderomotive}, and later generalised for time-modulated couplings in~\cite{qvarfort2019enhanced}, with the addition of linear and quadratic mechanical driving terms in~\cite{qvarfort2020time}. Crucially, the fully nonlinear optomechanical interaction allows for the preparation of non-Gaussian states, such as optical and mechanical cat-states. While many experiments today are successfully described by linearizing the optomechanical Hamiltonian (see~\cite{aspelmeyer2014cavity} and references therein), the nonlinear regime remains a key target.

All quantum systems inevitably couple to their surrounding environment, which can lead to dissipation, thermalization and decoherence of the system state. In the case of cavity optomechanical systems, there are two primary sources of noise: Firstly, dissipation that affects the radiation mode due to both internal and external losses, 
and secondly, mechanical thermal noise that affects the mechanical mode. The type of mechanical noise that arises is specific to the platform in question. For example, in optically levitated system, the noise on the mechanical mode arises mainly due to the trapping laser and gas collisions, while in clamped systems, vibrations and other disturbances dominate since they are directly transmitted to the system~\cite{aspelmeyer2014cavity}. 

There are in general two different approaches for modelling mechanical noise for optomechanical systems. In the linearized optomechanical regime, mechanical noise can be modelled and solve through a Fourier treatment of the quantum Langevin equations~\cite{aspelmeyer2014cavity}. In the nonlinear regime, one must usually instead solve a quantum master equation to fully model the state. A solution of the Gorini-Kossakowski-Sudarshan-Lindblad equation~\cite{lindblad1976generators, gorini1976completely} equation for mechanical dephasing noise was presented in~\cite{bose1997preparation}, while position noise (which is well-suited for models of levitated systems~\cite{romero2011quantum}) was modelled with a stochastic master equation in~\cite{bassi2005towards}  and further extended to the high-temperature-limit in~\cite{bernad2006quest}.
In all preceding studies of mechanical noise in the nonlinear optomechanical regime, the environment was presumed to have an Ohmic, or Markovian spectrum, where the bath retains no memory of the interaction with the system~\cite{breuer2016colloquium}. Indeed, a Markovian noise model has been presumed sufficient for modelling both dissipation of the cavity mode and mechanical thermalization noise in most experiments to-date. However, measurements of a clamped membrane found that the noise obeys a strongly non-Markovian profile~\cite{groeblacher2015observation}. There may also be additional benefits to considering non-Markovian noise. In addition, a number of theoretical proposals have shown that, in the linearized optomechanical regime, access to a non-Markovian environment in optomechanical systems can bring benefits in terms of enhanced sideband cooling~\cite{triana2016ultrafast} as well as for sensing in the linearized optomechanical regime~\cite{zhang2017optomechanical}. An analytic solution for non-Markovian mechanical noise in the nonlinear optomechanical regime, however, has yet to be developed.

\begin{figure}[t]
\centering
 \includegraphics[width=1\linewidth]{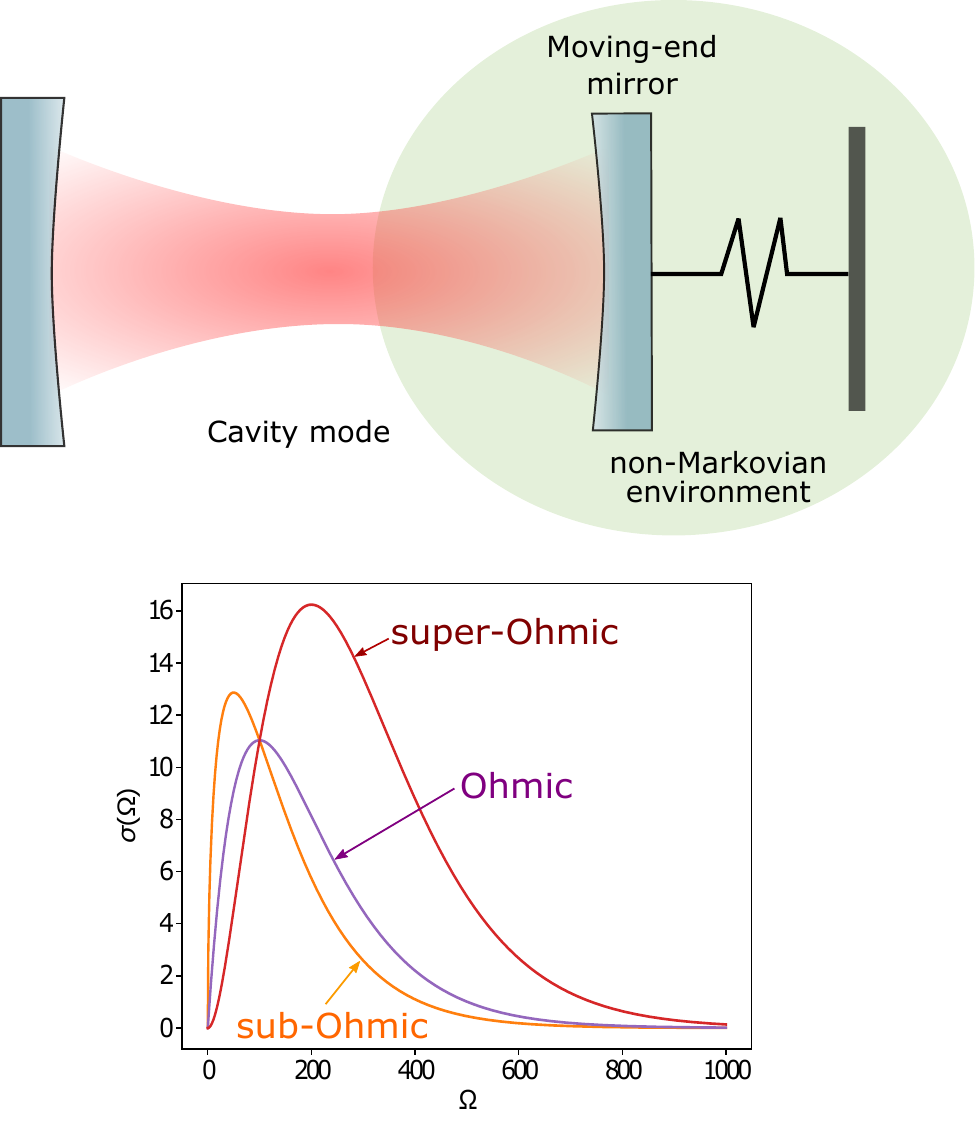}
\caption{Schematic figure of an example optomecahnical system in the form of a cavity with a moving-end mirror interacting with a non-Markovian environment. The spectrum of the environment displays a Markovian (Ohmic) or non-Markovian (sub-Ohmic or super-Ohmic) spectrum $\sigma(\Omega) =  \gamma \Omega \, (\Omega/\Omega_{C})^{k-1} e^{- \Omega/\Omega_C}$, where $\Omega$ is the frequency, $\Omega_C$ is the cutoff-frequency, here set to $\Omega_C/\omega_{\mathrm{m}}= 100$, and where $k$ is an integer number that denotes an Ohmic ($k= 1$), sub-Ohmic ($k= 1/2$), or super-Ohmic ($k = 2$) spectrum. }
\label{fig:system}
\end{figure}

In this work, we address this discrepancy by solving an exact model of a quantum optomechanical system in the nonlinear regime in the presence of non-Markovian mechanical noise (see~Figure~\ref{fig:system}). We consider a mechanical mode that interacts with a bath of quantum harmonic oscillators, which are collectively characterized by either a Markovian or non-Markovian spectrum. We derive an analytic solution to the dynamics by first solving the general quantum Langevin equation for the mechanical mode according to the Caldeira--Legget model~\cite{caldeira1983quantum}, which corresponds to modelling quantum Brownian motion. We then solve the evolution of the nonlinear optomechanical interaction term with a Lie algebra decoupling method~\cite{wei1963lie} (see~\cite{qvarfort2022solving} for a tutorial). By then tracing out the bath degrees-of-freedom, we obtain a fully analytic and exact solution to the system dynamics. The method developed here is distinct from the more commonly used Feynman--Vernon influence functional approach~\cite{feynman2000theory}, because we wait to trace out the bath modes until the very last step. Our analytic solution allows us to study a key property of optomechanical systems in the nonlinear regime, namely the strength of the self-Kerr optomechanical nonlinearity, which is key for the preparation of both intra-cavity and mechanical cat-states~\cite{bose1997preparation,mancini1997ponderomotive}.  We find that the influence of mechanical noise generally decreases the strength of the optomechanical nonlinearity, but that a strongly non-Markovian spectrum can enhance the nonlinearity, even beyond the values that can be achieved in a closed system. Our results suggest that engineering the spectrum of the environment could help strengthen the effects of the optomechanical nonlinearity. 

This work is structured as follows. In Section~\ref{sec:dynamics}, we introduce the optomechanical Hamiltonian and its coupling to the mechanical bath modes, then provide an exact solution for the resulting nonlinear dynamics. We then proceed in Section~\ref{sec:nonlinearity} to derive an expression for the optomechanical nonlinearity and study the influence of the bath. The work is concluded with some summarising remarks in Section~\ref{sec:conclusions}.

\section{Cavity optomechanical dynamics and quantum Brownian motion} \label{sec:dynamics}
Our goal is to derive an exact solution for the evolution of the cavity mode, mechanical mode, and the bath modes. We later trace out the bath modes to consider the strength of the optomechanical nonlinearity. 
Before we proceed, we here provide an overview of the procedure for deriving the results. We begin by 
(i) solving the quantum Langevin equation for the mechanical mode according the Caldeira-Legget model, which describes quantum Brownian motion. However, we do not yet trace out the bath modes, but keep them for the next part. Next, (ii) we consider the evolution of the nonlinear optomechanical interaction term in a frame that rotates with the mechanical mode and bath modes. We solve the  evolution of the interaction term in this picture using the Lie algebra decoupling method~\cite{wei1963lie}. Finally (iii), we write down the full time-evolution operator, which allows us to trace out the bath-degrees-of-freedom and consider the effect of the bath on the optomechanical nonlinearity. 

\subsection{Hamiltonian of the optomechanical system and bath} 

We start by considering the Hamiltonian of the cavity and mechanical modes, which are coupled together through radiation pressure. It reads
\begin{equation} \label{eq:optomechanical:Hamiltonian}
\hat H_{QOM} = \hat H_{0,\mathrm{c}} + \hat H_{0,\mathrm{m}} + \hat H_{I,\mathrm{cm}}, 
\end{equation}
where  $\hat H_{0,\mathrm{c}}$ is the free evolution of the cavity mode, $\hat H_{0,\mathrm{m}}$ is the free evolution of the mechanical mode, and where $\hat H_{I,\mathrm{cm}}$ describes the interaction. The terms are given by 
\begin{align}\label{eq:H0c:H0m:HIcm}
\hat H_{0,\mathrm{c}} &= \hbar \omega_{\mathrm{c}} \hat a^\dag \hat a,  \\
\hat H_{0,\mathrm{m}} &=  \frac{\hbar \omega_{\mathrm{m}} }{2} \hat P_m^2 + \frac{\hbar \omega_{\mathrm{m}} }{2} \hat X_m^2,  \label{eq:H0m}
 \\
\hat H_{I, \mathrm{cm}} &= - \hbar g(t) \, \hat a^\dag \hat a \,\bigl( \hat b^\dag + \hat b \bigr) \equiv - \sqrt{2} \hbar g \, \hat a^\dag \hat a \, \hat X_{\mathrm{m}}.  \label{eq:HIcm}
\end{align}
Here, $\omega_{\mathrm{c}}$ is the frequency of the cavity mode, $\omega_{\mathrm{m}}$ is the frequency of the mechanical mode, and $g(t)$ is the (possibly time-dependent) coupling constant that encodes the interaction strength between the cavity mode and mechanical mode. The operator $\hat a, \hat a^\dag$ and $\hat b, \hat b^\dag$ are bosonic annihilation and creation operators that describe the cavity and mechanical modes, respectively, satisfying $[\hat a, \hat a^\dag] = [\hat b, \hat b^\dag] = 1$. We have also defined the dimensionless quadrature operators $\hat X_{\mathrm{m}}$ and $\hat P_{\mathrm{m}}$ as
\begin{align}\label{eq:Xm:Pm:quads} 
&\hat X_{\mathrm{m}} = \frac{1}{\sqrt{2}}\left( \hat b^\dag + \hat b \right), 
&&\hat P_{\mathrm{m}} = \frac{i}{\sqrt{2}} \left( \hat b^\dag - \hat b \right). 
\end{align}
As can be seen from Eq.~\eqref{eq:HIcm}, the optomechanical interaction term can be written in terms of $\hat X_{\mathrm{m}}$. 
This later helps us incorporate the solution of the generalised quantum Langevin equation into the full nonlinear dynamics.

We then consider the case where the mechanical mode interacts with an external bath modelled as the degrees-of-freedom of an infinite collection of harmonic oscillators. We assume that the interaction between the mechanical mode and the bath is linear, which is captured by the Caldiera-Legget model where the mechanical mode is subjected to a fluctuating force~\cite{caldeira1983path}. The full Hamiltonian for the cavity mode, mechanical mode, and the bath therefore becomes
\begin{equation} \label{eq:full:Hamiltonian}
\hat H = \hat H_{0, \mathrm{c}} + \hat H_{0, \mathrm{m}} + \hat H_{0, \mathrm{B}} + \hat H_{I, \mathrm{cm}} + \hat H_{I, \mathrm{mB}}, 
\end{equation}
where $\hat H_{0,\mathrm{B}}$ is the free evolution of the bath and  $\hat H_{I, \mathrm{mB}}$ encodes the interaction between the mechanical mode and the bath. They are given by 
\begin{align}  
\hat H_{0, \mathrm{B}} &= \sum_{j=1}^N \frac{\hbar \Omega_j }{2} \left[\hat P_j^2 +  \hat X_j^2 \right], \label{eq:H0B}
\\
\hat H_{I, \mathrm{mB}} &= - \hbar \sum_{j=1}^N \kappa_j \hat X_m \hat X_j + \Delta V , \label{eq:HImB}
\end{align}
where we have defined the frequency $\Omega_j$ of the $j$th bath mode and the couplings $\kappa_j$ between the mechanics and the $j$th bath mode. Here,  $\Delta V$ is a counter-term added into the Hamiltonian to re-normalisate the potential to ensure that the frequencies stay the same~\cite{weiss2012quantum}. For a linearly coupled system, it is given by  
\begin{equation}
\Delta V = \hbar  \sum_{j = 1}^N  \frac{\kappa_j^2}{2 \Omega_j^2}\hat X_j^2 . 
\end{equation}
In Eq.~\eqref{eq:H0B}, we   also introduced the  dimensionless quadrature operators for the bath
\begin{align}\label{eq:Xj:Pj:quads}
&\hat X_j = \frac{1}{\sqrt{2}}\left( \hat c^\dag_j + \hat c_j \right), 
&&\hat P_j = \frac{i}{\sqrt{2}} \left( \hat c^\dag_j - \hat c_j \right). 
\end{align}
Our next step is to solve the dynamics of the mechanical subsystem and the bath.

\subsection{Brownian motion of the mechanical mode due to the bath} \label{sec:brownian:motion}
We proceed by considering the evolution of the mechanical mode under the influence of the bath modes. Our derivation follows those in~\cite{weiss2012quantum} and~\cite{gardiner2004quantum}. 	
Before we proceed, for simplicity we rescale all frequency-valued quantities by the mechanical resonant frequency $\omega_{\mathrm{m}}$. The quantities become $ \omega_{\mathrm{m}} t \rightarrow t$, $g(t) /\omega_{\mathrm{m}} \rightarrow g(t)$,  and $\kappa_j/\omega_{\mathrm{m}} \rightarrow \kappa_j$. Units of time and frequency can then be restored when necessary. In the derivation that follows, we also set $\hbar = 1$.

Our first task is to consider the evolution of the mechanical mode and the bath subsystems. We seek to derive a solution for the mechanical mode $\hat X_{\mathrm{m}}(t)$ as a function of its interaction with the bath modes. Later, we add back in the optomechanical interaction term. 
The Hamiltonian for the mechanical mode, the bath, and their interaction reads
\begin{equation} \label{eq:mechanics:bath:Hamiltonian}
\hat H_{\mathrm{mB}} =  \hat H_{0, \mathrm{m}} + \hat H_{0, \mathrm{B}} + \hat H_{I, \mathrm{mB}}, 
\end{equation}
where the three terms are given in Eq.~\eqref{eq:H0m},~\eqref{eq:H0B}, and~\eqref{eq:HImB}, respectively. 

To solve the dynamics induced by $\hat H_{\mathrm{mB}}$, we start by considering the equations of motion for $\hat X_{\mathrm{m}}(t)$ and $\hat X_j(t)$ in the Heisenberg picture. Note that we explicitly indicate the time-dependence of the evolved operator $\hat X_j(t)$, while $\hat X_{\mathrm{m}}$ or $\hat X_{\mathrm{m}}(0)$ refers to the initial quadrature operator. Similarly for the bath mode operators $\hat X_j(t)$. 
Under the Hamiltonian in Eq.~\eqref{eq:mechanics:bath:Hamiltonian}, $\hat X_{\mathrm{m}}$ and $\hat X_j$ evolve in the Heisenberg picture as 
\begin{align}
\ddot{\hat{X}}_{\mathrm{m}}(t) +  \hat X_{\mathrm{m}}(t) = \sum_{j=1}^N  \kappa_j \hat X_j(t),  \label{eq:eom:Xm} \\
\ddot{\hat{X}}_j(t) + \Omega_j^2 \hat X_j(t) = \kappa_j \Omega_j \hat X_{\mathrm{m}}(t), \label{eq:eom:Xj}
\end{align}
where we have used the fact that the canonical commutator relation for these dimensionless operators reads $[\hat X_{\mathrm{m}}, \hat P_{\mathrm{m}}] = i$. 

We proceed by solving the equation for the bath mode $\hat X_j(t)$ through the use of standard Green’s functions methods~\cite{weiss2012quantum}. We obtain the following solution for $\hat X_j(t)$
\begin{align} \label{eq:bath:solution}
\hat X_j (t) = \hat X_j ^{(0)} (t) + \kappa_j  \int ^t_0\mathrm{d}t' \,  \sin[\Omega_j (t - t')] \hat X_{\mathrm{m}} (t') , 
\end{align}
where we have defined  the free evolution of $\hat X_j(t)$ as 
\begin{equation}
\hat X_j^{(0)}(t) =  \frac{1}{\sqrt{2}}\left[ \hat c_j e^{- i \Omega_j t} + \hat c_j^\dag e^{i \Omega_j t} \right]. 
\end{equation}
Next, it is convenient to integrate Eq.~\eqref{eq:bath:solution} by parts, since a cosine term will be more favourable for future calculations~\cite{weiss2012quantum}. This allows us to write
\begin{align} \label{eq:intermediate:Xj:eom}
\hat X_j(t) &= \hat X_j ^{(0)} (t) +  \frac{\kappa_j}{\Omega_j} \left(\hat{X}_{\mathrm{m}}(t) - \cos(\Omega_j t) \hat X_{\mathrm{m}}(0) \right) \nonumber \\
&\quad - \kappa_j \int^t_0 \mathrm{d}t^\prime  \frac{\cos(\Omega_j (t - t^\prime))}{\Omega_j} \dot{\hat{X}}_{\mathrm{m}}(t^\prime). 
\end{align}
Here, the second term inside the brackets on the left-hand side can be understood as a slip term and a Lamb shift in the Hamiltonian~\cite{weiss2012quantum}. It can be shown that, by shifting the distribution of the thermal state of the bath using the counter-term $\Delta V$ that we added to the Hamiltonian (see Eq.~\eqref{eq:HImB}), this slip term can be removed. This transformation also ensures that the random force that affects the mechanical system displays proper statistical behavior. 

Neglecting the slip term (second term) in Eq.~\eqref{eq:intermediate:Xj:eom},  we write the equation for $\hat X_j(t)$ as 
\begin{align} \label{eq:bath:solution:cosine}
\hat X_j (t) &= \hat X_j ^{(0)} (t)   - \kappa_j \int^t_0 \mathrm{d}t^\prime  \frac{\cos(\Omega_j (t - t^\prime))}{\Omega_j} \dot{\hat{X}}_{\mathrm{m}}(t^\prime). 
\end{align}
We then proceed to insert Eq.~\eqref{eq:bath:solution:cosine} into the equation of motion for $\hat X_{\mathrm{m}}(t)$, shown in Eq.~\eqref{eq:eom:Xm}. The result is 
\begin{align} \label{eq:intermediate:Xm}
\ddot{\hat{X}}_{\mathrm{m}}(t) + \hat X_{\mathrm{m}}(t) &= \sum_{j= 1}^N \kappa_j^2 \frac{\hat{X}_{\mathrm{m}}(t)}{\Omega_j}    \\
&\quad -  \int^t_0 \mathrm{d}t^\prime  \Sigma(t - t^\prime)  \dot{\hat{X}}_{\mathrm{m}}(t^\prime) + \hat \xi(t) , \nonumber
\end{align}
where we have defined the memory kernel $\Sigma(t)$ as 
\begin{equation}
\Sigma(t) =  \sum_{j= 1}^N \frac{\kappa_j^2}{\Omega_j} \cos[\Omega_j(t - t^\prime)], 
\end{equation}
and where $\hat \xi(t)$ is a force term that encodes the interaction between the mechanical mode and the bath operators:
\begin{equation} \label{eq:definition:xi}
\hat \xi(t) = \sum_{j=1}^N \hat X_j^{(0)} = \frac{ 1}{\sqrt{2}} \sum_{j= 1}^N \kappa_j \left[ \hat c_j e^{- i \Omega_j t} + \hat c_j^\dag e^{i \Omega_j t} \right]. 
\end{equation}
We now assume that the spectrum of the bath is continuous. This means that we can make the following identification for the memory kernel by taking the continuum limit of the bath frequencies:
\begin{equation} \label{eq:memory:kernel}
\Sigma(t) = \sum_{j = 1}^N \frac{\kappa_j^2 }{\Omega_j}\cos(\Omega_j t) \rightarrow  \frac{2}{\pi} \int^\infty_0 \mathrm{d}\Omega \, \sigma(\Omega) \frac{\cos(\Omega t)}{\Omega}, 
\end{equation}
where $\sigma(\Omega)$ encodes the spectrum of the bath. The spectrum can be linked back to the coupling coefficients $\kappa_j$ through the relation
\begin{equation} \label{eq:spectrum:discrete:relation}
\sigma(\Omega) = \frac{\pi}{2} \sum_{j = 1}^N \kappa_j^2 \delta( \Omega - \Omega_j). 
\end{equation}
Inserting this into Eq.~\eqref{eq:intermediate:Xm}, we find that the equation of motion for $\hat X_{\mathrm{m}}(t)$ is given by 
\begin{align} \label{eq:generalised:langevin}
\ddot{\hat{X}}_{\mathrm{m}}(t) + \hat X_{\mathrm{m}}(t) +  \int^t_0 \mathrm{d}t^\prime  \Sigma(t - t^\prime)  \dot{\hat{X}}_{\mathrm{m}}(t^\prime) =  \hat \xi(t) .
\end{align}
Once we have solved this equation, we fully understand the evolution of the mechanical subsystem. 

The solution to Eq.~\eqref{eq:generalised:langevin} can be obtained by use of a Laplace- transformation. We define the following quantities, where each tilde denotes a Laplace-transformed quantity
\begin{align}\label{eq:Laplace:transforms}
\tilde{\hat{X}}_{\mathrm{m}}(s)  &= \int^\infty_0 \mathrm{d}t \, e^{- s t} \hat X_{\mathrm{m}}(t),  \nonumber \\
\tilde{{\Sigma}}(s) &= \int^\infty_0 \mathrm{d}t \, e^{- st} {\Sigma}(t),   \\
\tilde{\hat{\xi}}(s) &= \int^\infty_0 \mathrm{d}t \, e^{- st} \hat{\xi}(t) . \nonumber
\end{align}
Here, $s$ is a complex parameter with units of frequency. The Laplace transform of the memory kernel can also be written as
\begin{align} \label{eq:memory:kernel}
\tilde{{\Sigma}}(s) &= \int^\infty_0 \mathrm{d}t \, e^{- st} {\Sigma}(t) \nonumber \\
&=\frac{2}{\pi} \int^\infty_0 \mathrm{d}t \, e^{- st}  \int^\infty_0 \mathrm{d}\Omega \, \frac{ \sigma(\Omega)}{\Omega}\cos(\Omega t)  \nonumber \\
&=\frac{2}{\pi} \int^\infty_0 \mathrm{d}\Omega  \, \frac{ \sigma(\Omega)}{\Omega}\, \frac{s}{\Omega^2 + s^2}. 
\end{align}
Then, taking the Laplace transform of Eq.~\eqref{eq:generalised:langevin} and rearranging the resulting terms, we find the following solution for $\tilde{\hat{X}}_{\mathrm{m}}(s)$:
\begin{align} \label{eq:Xm:frequency:solution}
\tilde{\hat{X}}_{\mathrm{m}}(s) &=  g(s) \dot{\hat{X}}_{\mathrm{m}}(0) + \left(\frac{1 -  g(s)}{s} \right) \hat X_{\mathrm{m}}(0)  \nonumber \\
 &\quad+g(s) \tilde{\hat{\xi}}(s), 
\end{align}
where we have defined  the function
\begin{align} \label{eq:defined:gs}
g(s) = \frac{1}{s^2 + 1 +  s \tilde{\Sigma}(s) }. 
\end{align}
By then taking the inverse Laplace transform of Eq.~\eqref{eq:Xm:frequency:solution}, we recover the solution for $\hat X_{\mathrm{m}}(t)$ in the time-domain, which read
\begin{align} \label{eq:Xm:solution}
\hat X_{\mathrm{m}}(t) &= \hat X_{\mathrm{m}} \left( 1 -  \int^t_0 \mathrm{d}t’ G(t’) \right) +  \hat P_{\mathrm{m}} G(t) \nonumber \\
&\quad + \int^t_0 G(t - t') \hat \xi(t') \mathrm{d}t' . 
\end{align}
Here,  $G(t)$ is the inverse Laplace transform of $g(s)$, given by 
\begin{align} \label{eq:Greens:function}
G(t) = \int_{\mathbb{C}} \mathrm{d}s \frac{g(s)}{2 \pi i } e^{st}, 
\end{align}
where $\mathbb{C}$ indicates the Bromwich contour. 

We then use our expression for $\hat \xi(t')$ in Eq.~\eqref{eq:definition:xi} to rewrite Eq.~\eqref{eq:Xm:solution} in terms of the dimensionless quadrature operators of the bath modes. The exact solution reads
\begin{align} \label{eq:Xm:solution:clean}
\hat X_{\mathrm{m}}(t) &=  \alpha(t) \hat X_{\mathrm{m}}  + \beta(t) \hat P_{\mathrm{m}}  \nonumber \\
&\quad + \sum_j \kappa_j \left[ \alpha_j(t) \hat X_j + \beta_j(t) \hat P_j \right],   
\end{align}
where we have defined the following time-dependent and dimensionless functions
\begin{align} \label{eq:alpha:beta:coeffs}
\alpha(t) &= 1 -  \int^t_0 \mathrm{d}t^\prime \, G(t^\prime), \nonumber  \\
\beta(t) &= G(t),  
\end{align}
as well as
\begin{align} \label{eq:alphaj:betaj:coeffs}
\alpha_j(t) &= \int^t_0 \mathrm{d}t' \, G(t - t') \cos(\Omega_j t') ,  \nonumber\\
\beta_j(t) &= \int^t_0 \mathrm{d}t' \, G(t - t') \sin(\Omega_j t').
\end{align} 
The coefficients $\alpha(t)$, $\beta(t)$, $\alpha_j(t)$ and $\beta_j(t)$ completely characterise the evolution of the mechanical subsystem due to the bath modes. 
From Eq.~\eqref{eq:alpha:beta:coeffs}, we see that $G(t)$ is the key quantity that determine the response of the mechanical system to the bath. To calculate $G(t)$, we need an explicit form of the spectrum $\sigma(\Omega)$, which we give later in Eq.~\eqref{eq:spectrum}. 

Before moving on, we note that when there is no coupling between the bath and the mechanical mode, we find that $G(t) = \sin(t)$, which means that Eq.~\eqref{eq:Xm:solution:clean} simplifies to 
\begin{align} \label{eq:pure:Xm}
\hat X_{\mathrm{m}}(t) = \cos( t) \hat X_{\mathrm{m}} + \sin( t) \hat P_{\mathrm{m}}, 
\end{align}
which we recognise as the free evolution of the mechanical quadrature $\hat X_{\mathrm{m}}$.

\subsection{Dynamics of the optomechanical interaction term} 
We have solved the dynamics of the mechanical subsystem due to the influence of the bath modes, which resulted in Eq.~\eqref{eq:Xm:solution:clean}. Our next task is to incorporate the nonlinear optomechanical interaction term shown in Eq.~\eqref{eq:HIcm} into our solutions. 

To solve the full dynamics of the cavity mode, mechanical mode and the bath, we first define a frame that rotates with the time-evolution of the mechanical subsystem and the bath. This evolution is captured by the operator $\hat U_{\mathrm{m,B}}(t)$, which is given by 
\begin{align} \label{eq:UmB}
\hat U_{\mathrm{mB}}(t)  = \mathcal{T} \mathrm{exp} \left[ - i \int^t_0 \mathrm{d}t^\prime \, \left( \hat H_{\mathrm{0,m}} + \hat H_{\mathrm{0,B}} + \hat H_{I, \mathrm{mB}} \right) \right], 
\end{align}
where $\mathcal{T}$ denotes time-ordering, and where the Hamiltonian terms $\hat H_{\mathrm{0,m}}$, $\hat H_{\mathrm{0,B}}$ and $\hat H_{I, \mathrm{mB}}$ are defined in Eqs.~\eqref{eq:H0m},~\eqref{eq:H0B}, and   and~\eqref{eq:HImB}, respectively.

In the frame that rotates with the bath and the mechanical mode, the optomechanical interaction Hamiltonian term $\hat H_{I, \mathrm{cm}}$, defined in Eq.~\eqref{eq:HIcm}, evolves due to  $\hat U_{\mathrm{mB}}(t)$ as 
\begin{align}
 \hat H_{I, \rm{cm}}'(t) &=  \hat U_{\mathrm{mB}}^\dag (t) \, \hat H_{I, \rm{cm}} \, \hat U_{\rm{mB}}(t)  
\nonumber \\
&= -  \sqrt{2} \, \hbar g(t) \,  \hat a^\dag \hat a \hat X_{\mathrm{m}}(t), 
\end{align}
where $\hat H_{I, \rm{cm}}'(t) $ denotes the interaction term evolving in the frame of the bath and the mechanical mode, and where $\hat X_{\mathrm{m}}(t)$ is the evolution of the mechanical mode due to the bath, which we previously derived in Section~\ref{sec:brownian:motion}. It is given in Eq.~\eqref{eq:Xm:solution:clean}. We therefore find that the interaction Hamiltonian in the bath frame as 
\begin{align} \label{eq:int:pic:HIcm}
\hat H_{I, \mathrm{cm}}'(t)
&=- \sqrt{2} \, \hbar g(t) \, \hat a^\dag \hat a \bigl[\alpha(t) \hat X_{\mathrm{m}} + \beta(t) \hat P_{\mathrm{m}}  \nonumber \\
&\qquad\quad  + \kappa_j \sum_{j=1}^N \left( \alpha_j(t) \hat X_j + \beta_j(t) \hat P_j \right)\bigr]. 
\end{align}
where the coefficients $\alpha(t)$, $\beta(t)$, $\alpha_j(t)$ and $\beta_j(t)$ are given in Eqs~\eqref{eq:alpha:beta:coeffs} and~\eqref{eq:alphaj:betaj:coeffs}, respectively. 

Now that we know how $\hat H_{I, \mathrm{cm}}'(t)$ evolves due to the mechanical mode and the bath subsystem dynamics, our goal is to obtain a closed-form expression for its evolution $\hat U_{I,\rm{ cm}}(t)$, which is given by
\begin{align} \label{eq:UIcm:exact}
\hat U_{I,\rm{cm}}(t) = \mathcal{T}\mathrm{exp} \left[ - i \int^t_0 \mathrm{d}t' \, \hat H_{I, \mathrm{cm}}'(t) \right]. 
\end{align}
To find a solution to $\hat U_{I, \rm{cm}}$, we note that the operators in Eq.~\eqref{eq:int:pic:HIcm} form a finite-dimensional Lie algebra\footnote{We see this by taking the commutator of the operators and finding that the results commute with all other operators.}. This observation allows us to use a Lie algebra decoupling method, which provides a recipe for how to solve the resulting dynamics~\cite{wei1963lie}. 

By taking the commutator of the operators in Eq.~\eqref{eq:int:pic:HIcm}, we find that the Lie algebra that generate the evolution of $\hat U_{\mathrm{cm}}(t)$ are given by 
\begin{align} \label{eq:system:bath:algebra}
&\hat a^\dag \hat a\,  \hat X_{\mathrm{m}}, && \hat a^\dag \hat a \hat \, P_{\mathrm{m}}, & \nonumber \\
& \hat a^\dag \hat a\, \hat X_j, && \hat a^\dag \hat a \, \hat P_j, && (\hat a^\dag \hat a )^2,
\end{align}
where $\hat X_{\mathrm{m}}, \hat P_{\mathrm{m}}, \hat X_j,$ and $\hat P_j$ are defined in Eqs.~\eqref{eq:Xm:Pm:quads} and~\eqref{eq:Xj:Pj:quads}, and where the self-Kerr term $(\hat a^\dag \hat a)^2$ arises due to taking the commutator of $\hat a^\dag \hat a \hat X_{\mathrm{m}}$ and $\hat a^\dag \hat a \hat P_{\mathrm{m}}$, as well as $\hat a^\dag \hat a \hat X_j$ and $\hat a^\dag \hat a \hat P_j$. From these algebra elements, we note that the bath modes couple directly to the cavity mode. The interaction has been transduced through the optomechanical interaction term, and we later find that it has a direct impact on the strength of the optomechanical nonlinearity. 

Having identified the algebra elements in Eq.~\eqref{eq:system:bath:algebra}, we state the  ansatz for the time-evolution of the optomechanical interaction term $\hat U_{I, \mathrm{cm}}(t)$ in the frame of the bath and the mechanical mode as 
\begin{align} \label{eq:ansatz:nonlinear:U}
\hat U_{I, \rm{cm}}(t)&= e^{-i\,F_{a}\,(\hat a^\dag \hat a)^2}\,  e^{-i\,F_{X}\,\hat a^\dag \hat a\,\hat X_{\mathrm{m}}}
\,e^{-i\,F_{P}\,\hat a^\dag \hat a\,\hat P_{\mathrm{m}}} \,\nonumber \\
&\quad \times \prod_{j = 1}^N e^{- i \, F_{j, X} \, \hat a^\dag \hat a\, \hat X_j} \, e^{- i \, F_{j, P} \, \hat a^\dag \hat a \, \hat P_j}, 
\end{align}
where we have introduced the real time-dependent  coefficients $F_a, $ $F_X$, $F_P$, $F_{j,X}$, and  $F_{j,P}$, which all depend on time $t$, which we have suppressed to keep the notation simple. 

Our goal is now to obtain a set of differential equations for the $F$ coefficients. If we can solve these equations, we have obtained a full solution for the optomechanical dynamics. To do so, we follow the method outlined in~\cite{wei1963lie}. The first step is to note that differentiating the time-evolution operator $\hat U_{I,\rm{cm}}(t)$ in  Eq.~\eqref{eq:UIcm:exact} gives us 
\begin{align}
\frac{d}{dt} \hat U_{I,\rm{cm}}(t) = - i \hat H_{I,\rm{cm}}'(t) \hat U_{I,\rm{cm}}(t) . 
\end{align}
We proceed by similarly differentiating the ansatz in Eq.~\eqref{eq:ansatz:nonlinear:U} with respect to time $t$ and multiplying the result by $\hat U_{I,\rm{cm}}^\dag (t)$ from the right. Through use of the chain rule, we find 
\begin{align}  \label{eq:first:derivative:step}
&\left[ \frac{d}{dt} \hat U_{I,\rm{cm}}(t)\right] \hat U_{I,\rm{cm}}^\dag(t)= - i \dot{F}_a (\hat a^\dag \hat a)^2  - i \dot{F}_X \hat a^\dag \hat a \hat X_m \nonumber \\
&- i \dot{F}_P \,  \hat a^\dag \hat a \, e^{- i F_X \hat a^\dag \hat a \hat X_m} \,  \hat P_m \, e^{i F_X \hat a^\dag \hat a \hat X_m}   \\
&- i \hat a^\dag \hat a \sum_{j = 1}^N \left( \dot{F}_{j,X} \hat X_j + \dot{F}_{j, P} e^{- i F_{j,X} \hat a^\dag \hat a\hat X_j} \,  \hat P_m \, e^{i F_{j, X} \hat a^\dag \hat a \hat X_j} \right). \nonumber
\end{align}
To simplify Eq.~\eqref{eq:first:derivative:step} further, we make use of the relation
\begin{align}
e^{-i \theta \hat X_m} \hat P_m e^{ i \theta \hat X_m} = \hat P_m + \theta, 
\end{align}
which allows us to write Eq.~\eqref{eq:first:derivative:step} as 
\begin{align} 
&\left( \frac{d}{dt} \hat U_{I,\rm{cm}}(t)\right) \hat U_{I,\rm{cm}}^\dag(t)= - i \dot{F}_a (\hat a^\dag \hat a)^2  - i \dot{F}_X \hat a^\dag \hat a \hat X_m \nonumber \\
&\quad - i \dot{F}_P \,  \hat a^\dag \hat a \left(  \hat P_m +  F_X \hat a^\dag \hat a\right) \,   \\
&\quad- i  \sum_{j = 1}^N \left( \dot{F}_{j,X} \hat a^\dag \hat a \hat X_j + \dot{F}_{j,P}  \, \hat a^\dag \hat a \left(  \hat P_j + F_{j,X} \hat a^\dag \hat a \right) \,  \right].\nonumber
\end{align}
By then equating this expression with $\hat H_{\rm{I,cm}}(t')$ in Eq.~\eqref{eq:int:pic:HIcm} and removing the  factors of $-i$, we find 
\begin{align}
&- \sqrt{2} \hbar g(t) \hat a^\dag \hat a \bigl[\alpha(t) \hat X_{\mathrm{m}} + \beta(t) \hat P_{\mathrm{m}}  \nonumber \\
&\qquad + \kappa_j \sum_{j=1}^N \left( \alpha_j(t) \hat X_j + \beta_j(t) \hat P_j \right)\bigr] \nonumber \\
&=   \dot{F}_a (\hat a^\dag \hat a)^2  + \dot{F}_X \hat a^\dag \hat a \hat X_m +  \dot{F}_P \,  \hat a^\dag \hat a \left(  \hat P_m +  F_X \hat a^\dag \hat a\right) \, \nonumber  \\
&\quad +  \sum_{j = 1}^N \left( \dot{F}_{j,X} \hat a^\dag \hat a \hat X_j + \dot{F}_{j,P}  \, \hat a^\dag \hat a \left(  \hat P_j + F_{j,X} \hat a^\dag \hat a \right) \,  \right].
\end{align}
By invoking the linear independence of the operators, we  identify the following differential equations for the $F$ coefficients
\begin{align} \label{eq:diff:eqs}
&\dot{F}_X = - \sqrt{2}g(t) \, \alpha(t) , &&
\dot{F}_P = -\sqrt{2} g(t) \, \beta(t), \nonumber \\
&\dot{F}_{j,X} = - \sqrt{2}g(t) \, \sum_{j = 1}^N \alpha_j(t), &&
\dot{F}_{j,P} = - \sqrt{2}g(t) \, \sum_{j = 1}^N \beta_j(t), \nonumber \\
&\dot{F}_a  = -  \dot{F}_P F_X -  \dot{F}_{j,P} F_{j,X}  .
\end{align}
Here, the last equation for $F_a$ follows from the fact that $(\hat a^\dag \hat a)^2$ does not appear in the optomechanical Hamiltonian, but is rather generated by the dynamics. 

By then integrating the expressions in Eq.~\eqref{eq:diff:eqs}, we find 
\begin{align} \label{eq:F:coefficients}
F_{a}(t) &=  - 2 \int^t_0 \mathrm{d}t' \, g(t') \, \beta(t') \int^{t'}_0 \mathrm{d}t'' \,  g(t'') \, \alpha(t'') \nonumber \\
&\hspace{-0.5cm} - 2 \sum_{j = 1}^N \kappa_j^2  \int^t_0 \mathrm{d}t' \, g(t') \, \beta _j(t') \int^{t'}_0 \mathrm{d}t'' \, g(t'') \,\alpha_j(t'') , \nonumber \\
F_{X}(t) &= -\sqrt{2} \int^t_0 \mathrm{d}t' \, g(t') \,\alpha(t'), \nonumber \\
F_{P}(t) &= -\sqrt{2} \int^t_0 \mathrm{d} t' \, g(t') \, \beta (t') , \nonumber \\
F_{j,X}(t) &= - \sqrt{2}\kappa_j\int^t_0 \mathrm{d} t' \,  g(t') \,  \alpha_j(t') , \nonumber \\
F_{j,P}(t) &=- \sqrt{2}\kappa_j \int^t_0 \mathrm{d}t' \, g(t') \,  \beta_j(t'), 
\end{align}
where we recall that $\kappa_j$ is the coupling between the bath and the mechanical modes, and where $\alpha(t)$, $\beta(t)$, $\alpha_j(t)$, and $\beta_j(t)$ are given in Eqs.~\eqref{eq:alpha:beta:coeffs} and~\eqref{eq:alphaj:betaj:coeffs}.

We now make a two observations based on the form of $\hat U_{I,\mathrm{cm}}(t)$ in Eq.~\eqref{eq:ansatz:nonlinear:U} and the coefficients in Eq.~\eqref{eq:F:coefficients}: 
(i) The optomechanical interaction term leads to entanglement between the bath modes and the cavity mode (as evidenced by the appearance of $\hat a^\dag\hat a\, \hat X_j$ and $\hat a^\dag \hat a\,\hat P_j$). The coupling to the bath provides an effective shift of the self-Kerr nonlinearity, which is characterised by the coefficient $F_a(t)$. These observations align with previous results, which showed that noise on the cavity and mechanical modes in a strongly coupled optomechanical system cannot be treated separately~\cite{hu2015quantum}. 
(ii) All coefficients in Eq.~\eqref{eq:F:coefficients} are proportional to the optomechanical coupling strength $g(t)$ because the interaction between the cavity mode and the bath is mediated through the mechanical mode. The stronger the coupling is, the bigger the influence of the bath of the cavity mode.

Now, we note that $F_a(t)$ is made up of two terms: the first, which arises from the unitary dynamics, and the second, which contains the bath coupling $\kappa_j$. We now invoke the assumption that the infinitely many bath modes form a continuum in terms of frequencies and couplings. The identity which then allows us to relate the bath couplings to the spectrum reads
\begin{equation} \label{eq:continuous:spectrum}
\sum_{j = 1}^N \kappa_j^2 =\int^\infty_0 \mathrm{d}\Omega  \, \sigma(\Omega). 
\end{equation}
With in mind, $F_a$ can be written as 
\begin{align} \label{eq:def:Fa}
&F_a(t) =-2 \int^t_0 \mathrm{d}t' \, g(t') \, \beta(t') \int^{t'}_0 \mathrm{d}t'' \,  g(t'') \, \alpha(t'')  \\
& -2\int_0^\infty \mathrm{d}\Omega \,   \sigma(\Omega)  \int^t_0 \mathrm{d}t' \, g(t') \, \bar{\beta}(t') \int^{t'}_0 \mathrm{d}t'' \, g(t'') \,\bar{\alpha}(t''). \nonumber
\end{align}
where we  have defined the continuous analogues of $\alpha_j(t)$ and $\beta_j(t)$ as 
\begin{align} \label{eq:cont:alpha:beta}
\bar{\alpha}(t) &= \int^t_0 \mathrm{d}t' \, G(t - t') \cos(\Omega t') ,  \nonumber \\
\bar{\beta}(t) &= \int^t_0 \mathrm{d}t' \, G(t - t') \sin(\Omega t'). 
\end{align}
Note that we cannot yet invoke the same identity for $F_{j,X}(t)$ and $F_{j,P}(t)$ because we only have linear expressions of $\kappa_j$. We must first examine properties of the system and derive expressions with $\kappa_j^2$ before we can invoke the identity in Eq.~\eqref{eq:continuous:spectrum}.

\subsection{Full evolution of the cavity mode, mechanical mode, and bath modes }

Now that we have obtained the expressions for the dynamical coefficients shown in Eq.~\eqref{eq:F:coefficients}, we can state the full and exact solution to the dynamics of the cavity, mechanical, and bath modes. The result is 
\begin{align} \label{eq:time:evolution:result}
\hat U(t) &= \hat U_{\mathrm{mB}}(t) \,  e^{-i\,F_{a}\,(\hat a^\dag \hat a)^2}\,  e^{-i\,F_{X}\,\hat a^\dag \hat a\,\hat X_{\mathrm{m}}}
\,e^{-i\,F_{P}\,\hat a^\dag \hat a\,\hat P_{\mathrm{m}}} \,\nonumber \\
&\quad \times \prod_{j = 1}^N e^{- i \, F_{j, X} \, \hat a^\dag \hat a\, \hat X_j} \, e^{- i \, F_{j, P} \, \hat a^\dag \hat a \, \hat P_j}, 
\end{align}
where the operators $\hat X_\mathrm{m}$, $\hat P_{\mathrm{m}}$, $\hat X_j$, and $\hat P_j$ are defined in Eqs.~\eqref{eq:Xm:Pm:quads} and~\eqref{eq:Xj:Pj:quads}, respectively, and where the $F$ coefficients are given in Eq.~\eqref{eq:F:coefficients}. This is one of the main results of this paper. The Caldeira-Leggett solution is contained in $\hat U_{\mathrm{m,B}}(t)$, while the remaining terms encode the interaction between the cavity mode and the mechanical mode, as well as the mechanical mode and the bath modes, respectively.

\section{Strength of the optomechanical nonlinearity} \label{sec:nonlinearity}
We have derived a solution to the dynamics of an optomechanical system in the nonlinear regime in the presence of non-Markovian mechanical noise. We are now ready to examine the influence of non-Markovian noise on the optomechanical nonlinearity.

\subsection{Defining the nonlinearity} 

To characterise the strength of the optomechanical nonlinearity, we first note that we can combine the exponentials that contain quadrature operators in Eq.~\eqref{eq:time:evolution:result} as follows:
\begin{align}
&e^{-i\,F_{X}\,\hat a^\dag \hat a\,\hat X_{\mathrm{m}}}
\,e^{-i\,F_{P}\,\hat a^\dag \hat a\,\hat P_{\mathrm{m}}} \nonumber \\
 &\qquad=   \hat D(\hat a^\dag \hat a  K_m)  \, e^{- i F_X F_P (\hat a^\dag \hat a)^2/2}, 
\end{align}
where $\hat D(\xi) = e^{ \xi \hat b^\dag - \xi^* \hat b } $ are Weyl displacement operators, and where we have defined $K_{\rm{m}} =  ( F_P - i F_X)/\sqrt{2}$. 

This allows us to write the full evolution operator $\hat U(t)$ in Eq.~\eqref{eq:time:evolution:result} as 
\begin{align} \label{eq:time:evolution:result:compact}
\hat U(t) &= \hat U_{\mathrm{mB}}(t) \,    \prod_{j = 1}^N e^{-i\,[F_{a} +  F_X F_P/2 + F_{j,X} F_{j,P}/2] \,(\hat a^\dag \hat a)^2}\,  \nonumber \\
&\qquad \times  \hat D(\hat a^\dag \hat a  K_m)  \hat D(\hat a^\dag \hat a  K_j), 
\end{align}
where we have similarly defined $K_j =  (F_{j,P} - i F_{j,X})/\sqrt{2}$. 

The exponent in Eq.~\eqref{eq:time:evolution:result:compact} contains a time-dependent factor multiplied by the self-Kerr nonlinearity $(\hat a^\dag \hat a)^2$. All other terms are either linear in $\hat a^\dag \hat a$ or do not depend on the cavity mode. We therefore define the strength of the optomechanical nonlinearity as 
\begin{align}
\eta = |F_a + F_X F_P/2 + F_{j,X} F_{j,P}/2|, 
\end{align}
where we have included the absolute value because the self-Kerr nonlinearity induces a phase of the intra-cavity state which only depends on the magnitude of $\eta$. 

We now assume, for simplicity, that the optomechanical coupling is constant with $g(t) \equiv g_0$. Then, inserting the expressions for the $F$ coefficients shown in Eq.~\eqref{eq:F:coefficients} and taking the continuum limit of the bath modes as per the identity in Eq.~\eqref{eq:continuous:spectrum}, we find that the optomechanical nonlinearity in the presence of non-Markovian noise is given by 
\begin{align} \label{eq:def:of:eta}
\eta(t) &=  g_0^2\biggl| 2 \int^t_0 \mathrm{d}t' \, \beta(t') \int^{t'}_0 \mathrm{d}t'' \, \alpha(t'') \nonumber \\
&\quad -  \int^t_0 \mathrm{d}t'  \,\beta(t') \int^t_0 \mathrm{d}t'  \,\alpha(t'')  \\
&\quad +2   \,  \int^t_0 \mathrm{d}t' \,   \int^{t'}_0 \mathrm{d}t'' \, \int^\infty_0 \mathrm{d}\Omega \, \sigma(\Omega)\, \bar{\beta} (\Omega,t')\,\bar{\alpha}(\Omega,t'') \nonumber \\
&\quad- \int^t_0 \mathrm{d} t' \, \int^t_0 \mathrm{d} t'' \, \int^\infty_0 \mathrm{d}\Omega \, \sigma(\Omega) \,  \bar{\alpha}(\Omega,t') \bar{\beta}(\Omega,t'')  \biggr|,\nonumber
\end{align}
where we recall that $\alpha(t)$, $\beta(t)$, $\bar{\alpha}(\Omega,t)$, and $\bar{\beta}(\Omega,t)$ are given in Eqs.~\eqref{eq:alpha:beta:coeffs}  and~\eqref{eq:cont:alpha:beta}, and $\sigma(\Omega)$ is the spectrum of the bath, defined in Eq.~\eqref{eq:spectrum}. We also taken care to write out the dependence of $\Omega$ in $\bar{\alpha}(\Omega,t)$ and $\bar{\beta}(\Omega,t)$ explicitly. 

We can simplify Eq.~\eqref{eq:def:of:eta} further by noting that the second and fourth terms contain integrals over square regions, which can be divided into two triangular integrations. We find 
\begin{align}
&\eta(t) = g_0^2 \bigg| \int^t_0 \mathrm{d}t' \,  \int^{t'}_0 \mathrm{d}t'' \biggl\{ \,   \beta(t') \alpha(t'') -  \alpha(t')  \,\beta(t'')    \\
&\, + \int_0^\infty \mathrm{d}\Omega \, \sigma(\Omega)  \left[ \bar{\beta} (\Omega,t')  \,\bar{\alpha}(\Omega,t'') -  \bar{\alpha}(\Omega,t')\,  \bar{\beta}(\Omega,t'')  \right]  \bigg\} \biggr|. \nonumber
\end{align}
In general, $\eta(t)$ is challenging to compute because it contains multiple integrals. We recall from Eq.~\eqref{eq:alpha:beta:coeffs} that $\alpha(t)$, $\beta(t)$, $\bar{\alpha}(\Omega,t)$ and $\bar{\Omega,\beta}(t)$ also contains integrals over time, which results in us having to compute a five-dimensional integral (four over time, and one over $\Omega$). The multidimensional integrals arise because the effects from the environment are transduced into the cavity mode through the evolution of the mechanical mode. 

\subsection{Nonlinearity for closed dynamics} \label{sec:closed:dynamics}
In the absence of noise, sometimes also defined as the time for which the mechanical element undergoes coherent oscillations, we know from Eq.~\eqref{eq:pure:Xm} that (again considering a constant optomechanical coupling $g(t) \equiv g_0$, and momentarily restoring factors of $\omega_\mathrm{m}$)
\begin{align}
\alpha(t) = \cos(\omega_\mathrm{m} t), && \mbox{and} && \beta(t) = \frac{1}{\omega_{\mathrm{m}}}\sin(\omega_{\mathrm{m}} t). 
\end{align}
Inserting these expressions into the integrals in Eq.~\eqref{eq:F:coefficients} we find (with factors of $\omega_{\mathrm{m}}$ restored): 
\begin{align}
&\qquad F_a =-  \frac{g_0^2}{2\omega_{\mathrm{m}}^2} \left[ \omega_{\mathrm{m}} t - \cos{(\omega_\mathrm{m} t)} \sin{(\omega_\mathrm{m} t)} \right],    \\
&F_+ = - \frac{ \sqrt{2} g_0}{\omega_\mathrm{m}} \sin(\omega_\mathrm{m} t), \quad \quad 
F_- = -\frac{ \sqrt{2}g_0}{\omega_\mathrm{m}} \left[ 1 - \cos(\omega_\mathrm{m} t) \right]. \nonumber
\end{align}
These factors are equivalent to those previously found in the literature (see~\cite{qvarfort2019enhanced, qvarfort2020time, qvarfort2021master}), up to a factor of $\sqrt{2}$, which comes from our choice in this work of using the position and momentum quadratures as our Lie algebra basis. 

The remaining coefficients in Eq.~\eqref{eq:F:coefficients}, $F_{j,+}$ and $ F_{j,-}$ are zero, because there is no coupling to the environment. The expression for the optomechanical nonlinearity is then given by (as also found in~\cite{bose1997preparation}): 
\begin{align}
\eta(t) = \frac{g_0^2}{\omega_{\mathrm{m}}^2} \left[ \omega_{\mathrm{m}} t - \sin(\omega_{\mathrm{m}} t) \right]. 
\end{align}
We plot $\eta$ for closed dynamics with $\gamma = 0$ in Figure~\ref{fig:nonlinearity} (blue circles). We have chosen to set $g_0/\omega_{\mathrm{m}} = 1$ because it is merely a proportionality factor in $\eta(t)$ (in experiments, $g_0/\omega_{\mathrm{m}}$ is typically much smaller). The nonlinearity increases monotonically as a function of time with a period of $2\pi$. It corresponds to a phase accumulated by the cavity state, which allows for the preparation of cat-states~\cite{bose1997preparation, mancini1997ponderomotive}.

\subsection{Nonlinearity with mechanical noise} \label{sec:Markovian:spectrum}

\begin{figure*}
\centering
\subfloat[ \label{fig:low:cutoff}]{
  \includegraphics[width=0.4\linewidth,trim=1cm 0 -1cm 0]{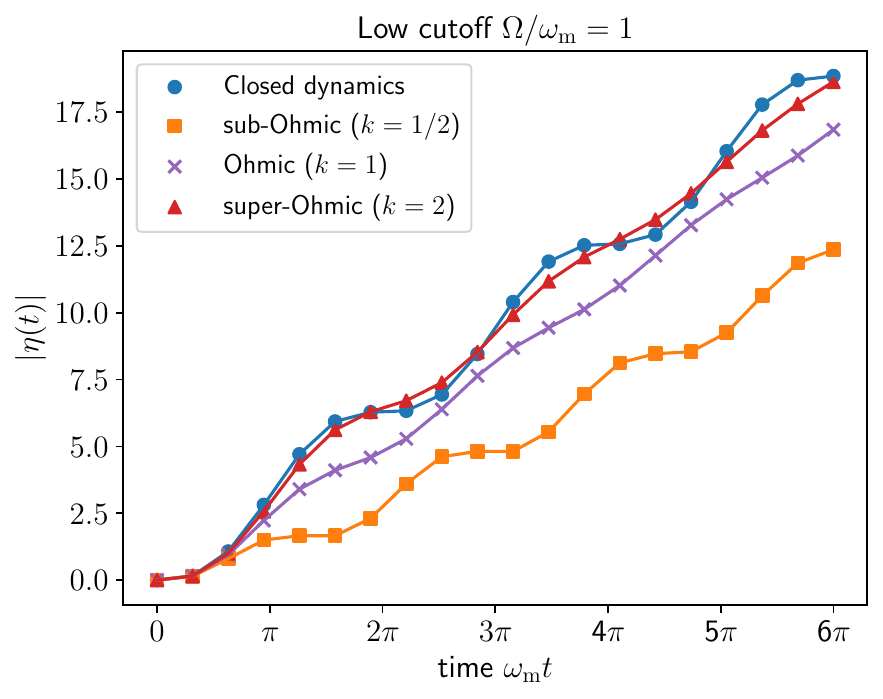}
}$\qquad $
\subfloat[ \label{fig:high:cutoff}]{
  \includegraphics[width=0.4\linewidth,trim=1cm 0 -1cm 0]{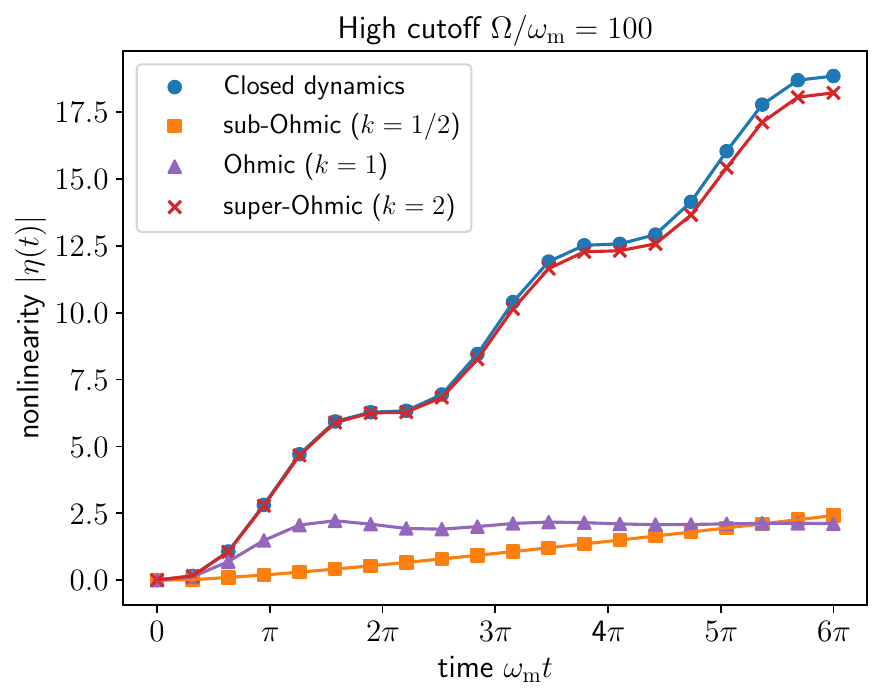}
}
\caption{Plot of the optomechanical nonlinearity $\eta(t)$ in Eq.~\eqref{eq:rewritten:eta} as a function of time for different bath spectra. (a) shows $\eta(t)$ for a frequency cutoff around the mechanical resonance frequency $\Omega_C/\omega_{\mathrm{m}} = 1$ and (b) shows $\eta(t)$ for $\Omega_C/\omega_{\mathrm{m}} = 100$. Both plots show the nonlinearity for unitary dynamics (blue circles), as well as for a sub-Ohmic (orange squares), Ohmic (purple triangles), and super-Ohmic (red crosses) spectrum. The dissipation rate for all cases is $\gamma/\omega_{\rm{m}} = 0.3$. We note that for both a low and high cutoff, a non-Markovian spectrum leads to an enhancement of the nonlinearity compared with the Marakovian case.  }
\label{fig:nonlinearity}
\end{figure*}

We now consider the strength of the nonlinearity in the presence of both Markovian (Ohmic) and non-Markovian mechanical noise. The general form of the spectrum is commonly given by~\cite{gardiner2004quantum,weiss2012quantum}
\begin{align}  \label{eq:spectrum}
\sigma(\Omega) =  \gamma \Omega \, (\Omega/\Omega_{C})^{k-1} e^{- \Omega/\Omega_C}, 
\end{align}
where $\gamma$ is the noise coefficient, $k$ is a positive number that determines the structure of the bath and $\Omega_C$ is the cutoff frequency. Here, different values of $k$ determine the shape of the spectrum. 
The Markovian spectrum is found by setting $k = 1$ in Eq.~\eqref{eq:spectrum} such that
$
\sigma(\Omega) = 
 \gamma \Omega \, e^{- \Omega/ \Omega_C} 
$,  where we recall that $\gamma$ is the dissipation rate and $\Omega_C$ is the cutoff. A Markovian bath is characterised by the fact that the bath retains no memory of the previous evolution of the system. 
In contrast, a non-Markovian bath with a sub-Ohmic ($k = 1/2$) or super-Ohmic ($k = 2$) spectrum corresponds to a can give rise to memory effects and back-action. 

To evaluate the strength of the nonlinearity $\eta(t)$, we must first integrate over the bath frequencies $\Omega$. By inserting the expressions for $\bar{\alpha}(\Omega,t)$ and $\bar{\beta}(\Omega,t)$ in Eq.~\eqref{eq:cont:alpha:beta}, we write $\eta(t)$ as 
\begin{align} \label{eq:rewritten:eta}
&\eta(t) = g_0^2 \biggl| \int^t_0 \mathrm{d}t' \,  \int^{t'}_0 \mathrm{d}t'' \biggl\{ \,  \left[ \beta(t') \alpha(t'') -  \alpha(t')  \,\beta(t'')  \right]   \nonumber \\
& +  \int^{t'}_0 \mathrm{d}t_\alpha \, \int^{t''}_0 \mathrm{d}t_\beta G(t' - t_\alpha)  \, G(t'' - t_\beta) C_k(t_\alpha - t_\beta)   \biggr\} \biggr|, 
\end{align}
where $G(t)$ is the Green's function in Eq.~\eqref{eq:Greens:function}, and where we have defined 
\begin{align}
C_k(t) = \int \mathrm{d}\Omega \,  \sigma(\Omega) \sin( \Omega t). 
\end{align}
For our specific choices of Markovian and non-Markovian spectra $C_k(t)$,  takes on analytic solutions. For $k = 1/2, 1, $ and $2$, we find 
\begin{align}
C_{1/2}(t) &= \frac{\sqrt{\pi } \gamma  \Omega_C \sin \left[\frac{3}{2} \tan^{-1}(t \Omega_C)\right]}{2 \left(t^2 \Omega_C^2+1\right)^{3/4}}, \nonumber \\
C_{1} (t) &= \frac{2 \gamma  t \Omega_C^2}{\left(t^2 \Omega_C^2+1\right)^2},  \nonumber \\
C_2 (t) &= -\frac{2 \gamma  t \Omega_C^2 \left(t^2 \Omega_C^2-3\right)}{\left(t^2 \Omega_C^2+1\right)^3}. 
\end{align}
In the limit where $\Omega_C \rightarrow \infty$, these expressions all tend to zero. As a result, the second term of $\eta(t)$ in Eq.~\eqref{eq:rewritten:eta} only contributes to the nonlinearity when the cutoff is small.

We proceed to evaluate $\eta(t)$ numerically for a sub-Ohmic ($k = 1/2$), an Ohmic ($k = 1$) and super-Ohmic ($k = 2$) spectrum for both small and large cutoffs. The results can be found in Figure~\ref{fig:nonlinearity}. We plot the nonlinearity as a function of time for a Markovian and non-Markovian spectrum for a low cutoff $\Omega/\omega_{\mathrm{m}} = 1$ in Figures~\ref{fig:low:cutoff}, and for a high cutoff $\Omega/\omega_{\mathrm{m}} = 100$ in Figure~\ref{fig:high:cutoff}. The dissipation rate has been set to $\gamma/\omega_{\mathrm{m}} = 0.3$ in both cases, and the optomechanical coupling strength to $g_0/\omega_{\mathrm{m}} = 1$, since it is merely a proportionality constant in $\eta(t)$. The plots show $\eta(t)$ for closed dynamics $\gamma = 0$ (blue circles), a sub-Ohmic spectrum ($k = 1/2$, orange squares), an Ohmic spectrum ($k = 1$, purple triangles) and a super-Ohmic spectrum ($k = 2$, red crosses).  

From the plots we see that the addition of mechanical noise generally reduces the strength of the optomechanical nonlinearity. Notably, the presence of Markovian noise consistently results in a lower value of $\eta(t)$ compared with closed dynamics. In the case of a high cutoff $\Omega /\omega_{\mathrm{m}} = 100$, the Markovian bath structure leads to a constant nonlinearity that stabilises to $\eta \sim 2.1$, presumably because the system reaches a steady-state which balances the noise and the influence of the optomechanical interaction term. 
In contrast, the inclusion of a non-Markovian spectrum appears largely beneficial, especially for the case of a super-Ohmic spectrum. Again for a large cutoff $\Omega /\omega_{\mathrm{m}} = 100$, we see that the super-Ohmic spectrum yields a nonlinearity that is almost unaffected by the noise. Indeed, even a sub-Ohmic spectrum does better than the Markovian case in the long-time limit. 

In general, a super-Ohmic spectrum yields the greatest enhancement to the optomechanical nonlinearity. Most notably, we see from Figure~\ref{fig:low:cutoff} that in the case of a low-cutoff $\Omega_C/\omega_{\mathrm{m}} = 1$, where the bath frequencies are truncated at the resonant frequency of the mechanical mode, the optomechanical nonlinearity is almost consistently as strong as for a closed system. That is, even a Markovian spectrum performs well for a low cutoff. Crucially, however, we see that a super-Ohmic spectrum can perform even better than the nonlinearity for a closed-system. This implies that by engineering the bath to have a low cutoff and a non-Markovian spectrum, it is possible to enhance the optomechanical nonlinearity beyond that for a closed system.

\section{Conclusion and outlook} \label{sec:conclusions}
We have derived a solution for the dynamics of an optomechanical systems in the nonlinear regime where the mechanical element experiences quantum Brownian motion by interacting with a bath of quantum harmonic oscillators. The solutions allow us to incorporate both Markovian and non-Markovian mechanical noise, which affects the dynamics and properties of the optomechanical system. 

To demonstrate the applicability of the model, we computed the strength of the optomechanical nonlinearity in the presence of Markovian and non-Markovian mechanical noise. We found that its value is generally reduced by the presence of mechanical noise, but that it is possible to mitigate the reduction and even enhance the nonlinearity by engineering the bath to have a highly non-Markovian spectrum. We also found, however, that for a Markovian noise bath and a large bath frequency cutoff, the optomechanical nonlinearity tends to a constant value. This potentially implies that the preparation of highly non-Gaussian states, such as cat-states, could be challenging in a Markovian environment. To fully determine the effects of non-Markovian noise, we would ultimately have to compare these results with the decoherence of off-diagonal density matrix elements of the cavity and mechanical modes. It is however challenging to consider the state of the mechanical mode within the formalism used here. The difficulty arises because tracing out the bath modes on their own is not straight-forward, as this requires a decoupling of the evolution operator of the bath and mechanical subsystems (see Eq.~\eqref{eq:UmB}). To consider the effects of non-Markovian noise on the mechanical state, we would either have to solve a fully non-Markovian master equation, such as that given by Caldeira and and Legget~\cite{caldeira1983path}, or perhaps a general non-Markovian master equation in Lindblad form in~\cite{zhang2012general}. In addition, we note that while we included the notation time-dependence in $g(t)$ in order to keep the solutions general, we did not consider the effects of time-modulated couplings and the interplay of non-Markovian dynamics in this work. We leave both of these investigations to future work.

\section*{Acknowledgments} 

I thank Igor Pikovski, Erik Aurell, Doug Plato, and David Edward Bruschi for fruitful discussions.  I also thank Stephen Stopyra for helpful advise on the numerical calculations. SQ is funded in part by the Wallenberg Initiative on Networks and Quantum Information (WINQ) and in part by the Marie Skłodowska–Curie Action IF programme “Nonlinear optomechanics for verification, utility, and sensing” (NOVUS) – Grant-Number 101027183. Nordita is partially supported by NordForsk.

\section*{Data availability statement} 
The code used to compute the optomechanical nonlinearity and produce the graphs in this work can be found in the following \href{https://github.com/sqvarfort/non-markov-optomechanics}{github repository}. The code was written in {\tt{Python}}~\cite{rossum2009python} and uses functions from the {\tt{SciPy}}~\cite{virtanen2020scipy} and {\tt{mpmath}}~\cite{mpmath} packages. 

\bibliographystyle{apsrev4-2}
\bibliography{non-markov-refs}

\begin{thebibliography}{40}%
\makeatletter
\providecommand \@ifxundefined [1]{%
 \@ifx{#1\undefined}
}%
\providecommand \@ifnum [1]{%
 \ifnum #1\expandafter \@firstoftwo
 \else \expandafter \@secondoftwo
 \fi
}%
\providecommand \@ifx [1]{%
 \ifx #1\expandafter \@firstoftwo
 \else \expandafter \@secondoftwo
 \fi
}%
\providecommand \natexlab [1]{#1}%
\providecommand \enquote  [1]{``#1''}%
\providecommand \bibnamefont  [1]{#1}%
\providecommand \bibfnamefont [1]{#1}%
\providecommand \citenamefont [1]{#1}%
\providecommand \href@noop [0]{\@secondoftwo}%
\providecommand \href [0]{\begingroup \@sanitize@url \@href}%
\providecommand \@href[1]{\@@startlink{#1}\@@href}%
\providecommand \@@href[1]{\endgroup#1\@@endlink}%
\providecommand \@sanitize@url [0]{\catcode `\\12\catcode `\$12\catcode
  `\&12\catcode `\#12\catcode `\^12\catcode `\_12\catcode `\%12\relax}%
\providecommand \@@startlink[1]{}%
\providecommand \@@endlink[0]{}%
\providecommand \url  [0]{\begingroup\@sanitize@url \@url }%
\providecommand \@url [1]{\endgroup\@href {#1}{\urlprefix }}%
\providecommand \urlprefix  [0]{URL }%
\providecommand \Eprint [0]{\href }%
\providecommand \doibase [0]{https://doi.org/}%
\providecommand \selectlanguage [0]{\@gobble}%
\providecommand \bibinfo  [0]{\@secondoftwo}%
\providecommand \bibfield  [0]{\@secondoftwo}%
\providecommand \translation [1]{[#1]}%
\providecommand \BibitemOpen [0]{}%
\providecommand \bibitemStop [0]{}%
\providecommand \bibitemNoStop [0]{.\EOS\space}%
\providecommand \EOS [0]{\spacefactor3000\relax}%
\providecommand \BibitemShut  [1]{\csname bibitem#1\endcsname}%
\let\auto@bib@innerbib\@empty
\bibitem [{\citenamefont {Aspelmeyer}\ \emph {et~al.}(2014)\citenamefont
  {Aspelmeyer}, \citenamefont {Kippenberg},\ and\ \citenamefont
  {Marquardt}}]{aspelmeyer2014cavity}%
  \BibitemOpen
  \bibfield  {author} {\bibinfo {author} {\bibfnamefont {M.}~\bibnamefont
  {Aspelmeyer}}, \bibinfo {author} {\bibfnamefont {T.~J.}\ \bibnamefont
  {Kippenberg}},\ and\ \bibinfo {author} {\bibfnamefont {F.}~\bibnamefont
  {Marquardt}},\ }\href {https://doi.org/10.1103/RevModPhys.86.1391} {\bibfield
   {journal} {\bibinfo  {journal} {Reviews of Modern Physics}\ }\textbf
  {\bibinfo {volume} {86}},\ \bibinfo {pages} {1391} (\bibinfo {year}
  {2014})}\BibitemShut {NoStop}%
\bibitem [{\citenamefont {Teufel}\ \emph {et~al.}(2011)\citenamefont {Teufel},
  \citenamefont {Donner}, \citenamefont {Li}, \citenamefont {Harlow},
  \citenamefont {Allman}, \citenamefont {Cicak}, \citenamefont {Sirois},
  \citenamefont {Whittaker}, \citenamefont {Lehnert},\ and\ \citenamefont
  {Simmonds}}]{teufel2011sideband}%
  \BibitemOpen
  \bibfield  {author} {\bibinfo {author} {\bibfnamefont {J.~D.}\ \bibnamefont
  {Teufel}}, \bibinfo {author} {\bibfnamefont {T.}~\bibnamefont {Donner}},
  \bibinfo {author} {\bibfnamefont {D.}~\bibnamefont {Li}}, \bibinfo {author}
  {\bibfnamefont {J.~W.}\ \bibnamefont {Harlow}}, \bibinfo {author}
  {\bibfnamefont {M.}~\bibnamefont {Allman}}, \bibinfo {author} {\bibfnamefont
  {K.}~\bibnamefont {Cicak}}, \bibinfo {author} {\bibfnamefont {A.~J.}\
  \bibnamefont {Sirois}}, \bibinfo {author} {\bibfnamefont {J.~D.}\
  \bibnamefont {Whittaker}}, \bibinfo {author} {\bibfnamefont {K.~W.}\
  \bibnamefont {Lehnert}},\ and\ \bibinfo {author} {\bibfnamefont {R.~W.}\
  \bibnamefont {Simmonds}},\ }\href {https://doi.org/10.1038/nature10261}
  {\bibfield  {journal} {\bibinfo  {journal} {Nature}\ }\textbf {\bibinfo
  {volume} {475}},\ \bibinfo {pages} {359} (\bibinfo {year}
  {2011})}\BibitemShut {NoStop}%
\bibitem [{\citenamefont {Chan}\ \emph {et~al.}(2011)\citenamefont {Chan},
  \citenamefont {Alegre}, \citenamefont {Safavi-Naeini}, \citenamefont {Hill},
  \citenamefont {Krause}, \citenamefont {Gr{\"o}blacher}, \citenamefont
  {Aspelmeyer},\ and\ \citenamefont {Painter}}]{chan2011laser}%
  \BibitemOpen
  \bibfield  {author} {\bibinfo {author} {\bibfnamefont {J.}~\bibnamefont
  {Chan}}, \bibinfo {author} {\bibfnamefont {T.~M.}\ \bibnamefont {Alegre}},
  \bibinfo {author} {\bibfnamefont {A.~H.}\ \bibnamefont {Safavi-Naeini}},
  \bibinfo {author} {\bibfnamefont {J.~T.}\ \bibnamefont {Hill}}, \bibinfo
  {author} {\bibfnamefont {A.}~\bibnamefont {Krause}}, \bibinfo {author}
  {\bibfnamefont {S.}~\bibnamefont {Gr{\"o}blacher}}, \bibinfo {author}
  {\bibfnamefont {M.}~\bibnamefont {Aspelmeyer}},\ and\ \bibinfo {author}
  {\bibfnamefont {O.}~\bibnamefont {Painter}},\ }\href
  {https://doi.org/10.1038/nature10461f} {\bibfield  {journal} {\bibinfo
  {journal} {Nature}\ }\textbf {\bibinfo {volume} {478}},\ \bibinfo {pages}
  {89} (\bibinfo {year} {2011})}\BibitemShut {NoStop}%
\bibitem [{\citenamefont {Deli{\'c}}\ \emph {et~al.}(2020)\citenamefont
  {Deli{\'c}}, \citenamefont {Reisenbauer}, \citenamefont {Dare}, \citenamefont
  {Grass}, \citenamefont {Vuleti{\'c}}, \citenamefont {Kiesel},\ and\
  \citenamefont {Aspelmeyer}}]{delic2020cooling}%
  \BibitemOpen
  \bibfield  {author} {\bibinfo {author} {\bibfnamefont {U.}~\bibnamefont
  {Deli{\'c}}}, \bibinfo {author} {\bibfnamefont {M.}~\bibnamefont
  {Reisenbauer}}, \bibinfo {author} {\bibfnamefont {K.}~\bibnamefont {Dare}},
  \bibinfo {author} {\bibfnamefont {D.}~\bibnamefont {Grass}}, \bibinfo
  {author} {\bibfnamefont {V.}~\bibnamefont {Vuleti{\'c}}}, \bibinfo {author}
  {\bibfnamefont {N.}~\bibnamefont {Kiesel}},\ and\ \bibinfo {author}
  {\bibfnamefont {M.}~\bibnamefont {Aspelmeyer}},\ }\href
  {https://doi.org/10.1126/science.aba3993} {\bibfield  {journal} {\bibinfo
  {journal} {Science}\ }\textbf {\bibinfo {volume} {367}},\ \bibinfo {pages}
  {892} (\bibinfo {year} {2020})}\BibitemShut {NoStop}%
\bibitem [{\citenamefont {Piotrowski}\ \emph {et~al.}(2023)\citenamefont
  {Piotrowski}, \citenamefont {Windey}, \citenamefont {Vijayan}, \citenamefont
  {Gonzalez-Ballestero}, \citenamefont {de~los R{\'\i}os~Sommer}, \citenamefont
  {Meyer}, \citenamefont {Quidant}, \citenamefont {Romero-Isart}, \citenamefont
  {Reimann},\ and\ \citenamefont {Novotny}}]{piotrowski2023simultaneous}%
  \BibitemOpen
  \bibfield  {author} {\bibinfo {author} {\bibfnamefont {J.}~\bibnamefont
  {Piotrowski}}, \bibinfo {author} {\bibfnamefont {D.}~\bibnamefont {Windey}},
  \bibinfo {author} {\bibfnamefont {J.}~\bibnamefont {Vijayan}}, \bibinfo
  {author} {\bibfnamefont {C.}~\bibnamefont {Gonzalez-Ballestero}}, \bibinfo
  {author} {\bibfnamefont {A.}~\bibnamefont {de~los R{\'\i}os~Sommer}},
  \bibinfo {author} {\bibfnamefont {N.}~\bibnamefont {Meyer}}, \bibinfo
  {author} {\bibfnamefont {R.}~\bibnamefont {Quidant}}, \bibinfo {author}
  {\bibfnamefont {O.}~\bibnamefont {Romero-Isart}}, \bibinfo {author}
  {\bibfnamefont {R.}~\bibnamefont {Reimann}},\ and\ \bibinfo {author}
  {\bibfnamefont {L.}~\bibnamefont {Novotny}},\ }\href
  {https://doi.org/10.1038/s41567-023-01956-1} {\bibfield  {journal} {\bibinfo
  {journal} {Nature Physics}\ ,\ \bibinfo {pages} {1}} (\bibinfo {year}
  {2023})}\BibitemShut {NoStop}%
\bibitem [{\citenamefont {Bild}\ \emph {et~al.}(2023)\citenamefont {Bild},
  \citenamefont {Fadel}, \citenamefont {Yang}, \citenamefont {von L{\"u}pke},
  \citenamefont {Martin}, \citenamefont {Bruno},\ and\ \citenamefont
  {Chu}}]{bild2023schrodinger}%
  \BibitemOpen
  \bibfield  {author} {\bibinfo {author} {\bibfnamefont {M.}~\bibnamefont
  {Bild}}, \bibinfo {author} {\bibfnamefont {M.}~\bibnamefont {Fadel}},
  \bibinfo {author} {\bibfnamefont {Y.}~\bibnamefont {Yang}}, \bibinfo {author}
  {\bibfnamefont {U.}~\bibnamefont {von L{\"u}pke}}, \bibinfo {author}
  {\bibfnamefont {P.}~\bibnamefont {Martin}}, \bibinfo {author} {\bibfnamefont
  {A.}~\bibnamefont {Bruno}},\ and\ \bibinfo {author} {\bibfnamefont
  {Y.}~\bibnamefont {Chu}},\ }\href
  {https://www.science.org/doi/full/10.1126/science.adf7553#:~:text=10.1126/science.adf7553}
  {\bibfield  {journal} {\bibinfo  {journal} {Science}\ }\textbf {\bibinfo
  {volume} {380}},\ \bibinfo {pages} {274} (\bibinfo {year}
  {2023})}\BibitemShut {NoStop}%
\bibitem [{\citenamefont {Qvarfort}\ \emph {et~al.}(2018)\citenamefont
  {Qvarfort}, \citenamefont {Serafini}, \citenamefont {Barker},\ and\
  \citenamefont {Bose}}]{qvarfort2018gravimetry}%
  \BibitemOpen
  \bibfield  {author} {\bibinfo {author} {\bibfnamefont {S.}~\bibnamefont
  {Qvarfort}}, \bibinfo {author} {\bibfnamefont {A.}~\bibnamefont {Serafini}},
  \bibinfo {author} {\bibfnamefont {P.~F.}\ \bibnamefont {Barker}},\ and\
  \bibinfo {author} {\bibfnamefont {S.}~\bibnamefont {Bose}},\ }\href
  {https://doi.org/10.1038/s41467-018-06037-z} {\bibfield  {journal} {\bibinfo
  {journal} {Nature Communications}\ }\textbf {\bibinfo {volume} {9}},\
  \bibinfo {pages} {3690} (\bibinfo {year} {2018})}\BibitemShut {NoStop}%
\bibitem [{\citenamefont {Schneiter}\ \emph {et~al.}(2020)\citenamefont
  {Schneiter}, \citenamefont {Qvarfort}, \citenamefont {Serafini},
  \citenamefont {Xuereb}, \citenamefont {Braun}, \citenamefont {R{\"a}tzel},\
  and\ \citenamefont {Bruschi}}]{schneiter2020optimal}%
  \BibitemOpen
  \bibfield  {author} {\bibinfo {author} {\bibfnamefont {F.}~\bibnamefont
  {Schneiter}}, \bibinfo {author} {\bibfnamefont {S.}~\bibnamefont {Qvarfort}},
  \bibinfo {author} {\bibfnamefont {A.}~\bibnamefont {Serafini}}, \bibinfo
  {author} {\bibfnamefont {A.}~\bibnamefont {Xuereb}}, \bibinfo {author}
  {\bibfnamefont {D.}~\bibnamefont {Braun}}, \bibinfo {author} {\bibfnamefont
  {D.}~\bibnamefont {R{\"a}tzel}},\ and\ \bibinfo {author} {\bibfnamefont
  {D.~E.}\ \bibnamefont {Bruschi}},\ }\href
  {https://doi.org/10.1103/PhysRevA.101.033834} {\bibfield  {journal} {\bibinfo
   {journal} {Physical Review A}\ }\textbf {\bibinfo {volume} {101}},\ \bibinfo
  {pages} {033834} (\bibinfo {year} {2020})}\BibitemShut {NoStop}%
\bibitem [{\citenamefont {Qvarfort}\ \emph
  {et~al.}(2021{\natexlab{a}})\citenamefont {Qvarfort}, \citenamefont {Plato},
  \citenamefont {Bruschi}, \citenamefont {Schneiter}, \citenamefont {Braun},
  \citenamefont {Serafini},\ and\ \citenamefont
  {R{\"a}tzel}}]{qvarfort2021optimal}%
  \BibitemOpen
  \bibfield  {author} {\bibinfo {author} {\bibfnamefont {S.}~\bibnamefont
  {Qvarfort}}, \bibinfo {author} {\bibfnamefont {A.~D.~K.}\ \bibnamefont
  {Plato}}, \bibinfo {author} {\bibfnamefont {D.~E.}\ \bibnamefont {Bruschi}},
  \bibinfo {author} {\bibfnamefont {F.}~\bibnamefont {Schneiter}}, \bibinfo
  {author} {\bibfnamefont {D.}~\bibnamefont {Braun}}, \bibinfo {author}
  {\bibfnamefont {A.}~\bibnamefont {Serafini}},\ and\ \bibinfo {author}
  {\bibfnamefont {D.}~\bibnamefont {R{\"a}tzel}},\ }\href
  {https://doi.org/10.1103/PhysRevResearch.3.013159} {\bibfield  {journal}
  {\bibinfo  {journal} {Physical Review Research}\ }\textbf {\bibinfo {volume}
  {3}},\ \bibinfo {pages} {013159} (\bibinfo {year}
  {2021}{\natexlab{a}})}\BibitemShut {NoStop}%
\bibitem [{\citenamefont {Rademacher}\ \emph {et~al.}(2020)\citenamefont
  {Rademacher}, \citenamefont {Millen},\ and\ \citenamefont
  {Li}}]{rademacher2020quantum}%
  \BibitemOpen
  \bibfield  {author} {\bibinfo {author} {\bibfnamefont {M.}~\bibnamefont
  {Rademacher}}, \bibinfo {author} {\bibfnamefont {J.}~\bibnamefont {Millen}},\
  and\ \bibinfo {author} {\bibfnamefont {Y.~L.}\ \bibnamefont {Li}},\ }\href
  {https://doi.org/10.1515/aot-2020-0019} {\bibfield  {journal} {\bibinfo
  {journal} {Advanced Optical Technologies}\ }\textbf {\bibinfo {volume} {9}},\
  \bibinfo {pages} {227} (\bibinfo {year} {2020})}\BibitemShut {NoStop}%
\bibitem [{\citenamefont {Bassi}\ \emph {et~al.}(2017)\citenamefont {Bassi},
  \citenamefont {Gro{\ss}ardt},\ and\ \citenamefont
  {Ulbricht}}]{bassi2017gravitational}%
  \BibitemOpen
  \bibfield  {author} {\bibinfo {author} {\bibfnamefont {A.}~\bibnamefont
  {Bassi}}, \bibinfo {author} {\bibfnamefont {A.}~\bibnamefont
  {Gro{\ss}ardt}},\ and\ \bibinfo {author} {\bibfnamefont {H.}~\bibnamefont
  {Ulbricht}},\ }\href {https://doi.org/10.1088/1361-6382/aa864f} {\bibfield
  {journal} {\bibinfo  {journal} {Classical and Quantum Gravity}\ }\textbf
  {\bibinfo {volume} {34}},\ \bibinfo {pages} {193002} (\bibinfo {year}
  {2017})}\BibitemShut {NoStop}%
\bibitem [{\citenamefont {Bose}\ \emph {et~al.}(2017)\citenamefont {Bose},
  \citenamefont {Mazumdar}, \citenamefont {Morley}, \citenamefont {Ulbricht},
  \citenamefont {Toro{\v{s}}}, \citenamefont {Paternostro}, \citenamefont
  {Geraci}, \citenamefont {Barker}, \citenamefont {Kim},\ and\ \citenamefont
  {Milburn}}]{bose2017spin}%
  \BibitemOpen
  \bibfield  {author} {\bibinfo {author} {\bibfnamefont {S.}~\bibnamefont
  {Bose}}, \bibinfo {author} {\bibfnamefont {A.}~\bibnamefont {Mazumdar}},
  \bibinfo {author} {\bibfnamefont {G.~W.}\ \bibnamefont {Morley}}, \bibinfo
  {author} {\bibfnamefont {H.}~\bibnamefont {Ulbricht}}, \bibinfo {author}
  {\bibfnamefont {M.}~\bibnamefont {Toro{\v{s}}}}, \bibinfo {author}
  {\bibfnamefont {M.}~\bibnamefont {Paternostro}}, \bibinfo {author}
  {\bibfnamefont {A.~A.}\ \bibnamefont {Geraci}}, \bibinfo {author}
  {\bibfnamefont {P.~F.}\ \bibnamefont {Barker}}, \bibinfo {author}
  {\bibfnamefont {M.}~\bibnamefont {Kim}},\ and\ \bibinfo {author}
  {\bibfnamefont {G.}~\bibnamefont {Milburn}},\ }\href
  {https://doi.org/10.1103/PhysRevLett.119.240401} {\bibfield  {journal}
  {\bibinfo  {journal} {Physical Review Letters}\ }\textbf {\bibinfo {volume}
  {119}},\ \bibinfo {pages} {240401} (\bibinfo {year} {2017})}\BibitemShut
  {NoStop}%
\bibitem [{\citenamefont {Marletto}\ and\ \citenamefont
  {Vedral}(2017)}]{marletto2017gravitationally}%
  \BibitemOpen
  \bibfield  {author} {\bibinfo {author} {\bibfnamefont {C.}~\bibnamefont
  {Marletto}}\ and\ \bibinfo {author} {\bibfnamefont {V.}~\bibnamefont
  {Vedral}},\ }\href {https://doi.org/10.1103/PhysRevLett.119.240402}
  {\bibfield  {journal} {\bibinfo  {journal} {Physical Review Letters}\
  }\textbf {\bibinfo {volume} {119}},\ \bibinfo {pages} {240402} (\bibinfo
  {year} {2017})}\BibitemShut {NoStop}%
\bibitem [{\citenamefont {Bose}\ \emph {et~al.}(1997)\citenamefont {Bose},
  \citenamefont {Jacobs},\ and\ \citenamefont {Knight}}]{bose1997preparation}%
  \BibitemOpen
  \bibfield  {author} {\bibinfo {author} {\bibfnamefont {S.}~\bibnamefont
  {Bose}}, \bibinfo {author} {\bibfnamefont {K.}~\bibnamefont {Jacobs}},\ and\
  \bibinfo {author} {\bibfnamefont {P.}~\bibnamefont {Knight}},\ }\href
  {https://doi.org/10.1103/PhysRevA.56.4175} {\bibfield  {journal} {\bibinfo
  {journal} {Physical Review A}\ }\textbf {\bibinfo {volume} {56}},\ \bibinfo
  {pages} {4175} (\bibinfo {year} {1997})}\BibitemShut {NoStop}%
\bibitem [{\citenamefont {Mancini}\ \emph {et~al.}(1997)\citenamefont
  {Mancini}, \citenamefont {Man'{k}o},\ and\ \citenamefont
  {Tombesi}}]{mancini1997ponderomotive}%
  \BibitemOpen
  \bibfield  {author} {\bibinfo {author} {\bibfnamefont {S.}~\bibnamefont
  {Mancini}}, \bibinfo {author} {\bibfnamefont {V.}~\bibnamefont {Man'{k}o}},\
  and\ \bibinfo {author} {\bibfnamefont {P.}~\bibnamefont {Tombesi}},\ }\href
  {https://doi.org/10.1103/PhysRevA.55.3042} {\bibfield  {journal} {\bibinfo
  {journal} {Physical Review A}\ }\textbf {\bibinfo {volume} {55}},\ \bibinfo
  {pages} {3042} (\bibinfo {year} {1997})}\BibitemShut {NoStop}%
\bibitem [{\citenamefont {Qvarfort}\ \emph {et~al.}(2019)\citenamefont
  {Qvarfort}, \citenamefont {Serafini}, \citenamefont {Xuereb}, \citenamefont
  {R{\"a}tzel},\ and\ \citenamefont {Bruschi}}]{qvarfort2019enhanced}%
  \BibitemOpen
  \bibfield  {author} {\bibinfo {author} {\bibfnamefont {S.}~\bibnamefont
  {Qvarfort}}, \bibinfo {author} {\bibfnamefont {A.}~\bibnamefont {Serafini}},
  \bibinfo {author} {\bibfnamefont {A.}~\bibnamefont {Xuereb}}, \bibinfo
  {author} {\bibfnamefont {D.}~\bibnamefont {R{\"a}tzel}},\ and\ \bibinfo
  {author} {\bibfnamefont {D.~E.}\ \bibnamefont {Bruschi}},\ }\href
  {https://doi.org/10.1088/1367-2630/ab1b9e} {\bibfield  {journal} {\bibinfo
  {journal} {New Journal of Physics}\ } (\bibinfo {year} {2019})}\BibitemShut
  {NoStop}%
\bibitem [{\citenamefont {Qvarfort}\ \emph {et~al.}(2020)\citenamefont
  {Qvarfort}, \citenamefont {Serafini}, \citenamefont {Xuereb}, \citenamefont
  {Braun}, \citenamefont {R{\"a}tzel},\ and\ \citenamefont
  {Bruschi}}]{qvarfort2020time}%
  \BibitemOpen
  \bibfield  {author} {\bibinfo {author} {\bibfnamefont {S.}~\bibnamefont
  {Qvarfort}}, \bibinfo {author} {\bibfnamefont {A.}~\bibnamefont {Serafini}},
  \bibinfo {author} {\bibfnamefont {A.}~\bibnamefont {Xuereb}}, \bibinfo
  {author} {\bibfnamefont {D.}~\bibnamefont {Braun}}, \bibinfo {author}
  {\bibfnamefont {D.}~\bibnamefont {R{\"a}tzel}},\ and\ \bibinfo {author}
  {\bibfnamefont {D.~E.}\ \bibnamefont {Bruschi}},\ }\href
  {https://doi.org/10.1088/1751-8121/ab64d5} {\bibfield  {journal} {\bibinfo
  {journal} {Journal of Physics A: Mathematical and Theoretical}\ }\textbf
  {\bibinfo {volume} {53}},\ \bibinfo {pages} {075304} (\bibinfo {year}
  {2020})}\BibitemShut {NoStop}%
\bibitem [{\citenamefont {Lindblad}(1976)}]{lindblad1976generators}%
  \BibitemOpen
  \bibfield  {author} {\bibinfo {author} {\bibfnamefont {G.}~\bibnamefont
  {Lindblad}},\ }\href {https://doi.org/10.1007/BF01608499} {\bibfield
  {journal} {\bibinfo  {journal} {Communications in Mathematical Physics}\
  }\textbf {\bibinfo {volume} {48}},\ \bibinfo {pages} {119} (\bibinfo {year}
  {1976})}\BibitemShut {NoStop}%
\bibitem [{\citenamefont {Gorini}\ \emph {et~al.}(1976)\citenamefont {Gorini},
  \citenamefont {Kossakowski},\ and\ \citenamefont
  {Sudarshan}}]{gorini1976completely}%
  \BibitemOpen
  \bibfield  {author} {\bibinfo {author} {\bibfnamefont {V.}~\bibnamefont
  {Gorini}}, \bibinfo {author} {\bibfnamefont {A.}~\bibnamefont
  {Kossakowski}},\ and\ \bibinfo {author} {\bibfnamefont {E.~C.~G.}\
  \bibnamefont {Sudarshan}},\ }\href {https://doi.org/10.1063/1.522979}
  {\bibfield  {journal} {\bibinfo  {journal} {Journal of Mathematical Physics}\
  }\textbf {\bibinfo {volume} {17}},\ \bibinfo {pages} {821} (\bibinfo {year}
  {1976})}\BibitemShut {NoStop}%
\bibitem [{\citenamefont {Romero-Isart}(2011)}]{romero2011quantum}%
  \BibitemOpen
  \bibfield  {author} {\bibinfo {author} {\bibfnamefont {O.}~\bibnamefont
  {Romero-Isart}},\ }\href {https://doi.org/10.1103/PhysRevA.84.052121}
  {\bibfield  {journal} {\bibinfo  {journal} {Physical Review A}\ }\textbf
  {\bibinfo {volume} {84}},\ \bibinfo {pages} {052121} (\bibinfo {year}
  {2011})}\BibitemShut {NoStop}%
\bibitem [{\citenamefont {Bassi}\ \emph {et~al.}(2005)\citenamefont {Bassi},
  \citenamefont {Ippoliti},\ and\ \citenamefont {Adler}}]{bassi2005towards}%
  \BibitemOpen
  \bibfield  {author} {\bibinfo {author} {\bibfnamefont {A.}~\bibnamefont
  {Bassi}}, \bibinfo {author} {\bibfnamefont {E.}~\bibnamefont {Ippoliti}},\
  and\ \bibinfo {author} {\bibfnamefont {S.~L.}\ \bibnamefont {Adler}},\ }\href
  {https://doi.org/10.1103/PhysRevLett.94.030401} {\bibfield  {journal}
  {\bibinfo  {journal} {Physical Review Letters}\ }\textbf {\bibinfo {volume}
  {94}},\ \bibinfo {pages} {030401} (\bibinfo {year} {2005})}\BibitemShut
  {NoStop}%
\bibitem [{\citenamefont {Bern{\'a}d}\ \emph {et~al.}(2006)\citenamefont
  {Bern{\'a}d}, \citenamefont {Di{\'o}si},\ and\ \citenamefont
  {Geszti}}]{bernad2006quest}%
  \BibitemOpen
  \bibfield  {author} {\bibinfo {author} {\bibfnamefont {J.~Z.}\ \bibnamefont
  {Bern{\'a}d}}, \bibinfo {author} {\bibfnamefont {L.}~\bibnamefont
  {Di{\'o}si}},\ and\ \bibinfo {author} {\bibfnamefont {T.}~\bibnamefont
  {Geszti}},\ }\href {https://doi.org/10.1103/PhysRevLett.97.250404} {\bibfield
   {journal} {\bibinfo  {journal} {Physical Review Letters}\ }\textbf {\bibinfo
  {volume} {97}},\ \bibinfo {pages} {250404} (\bibinfo {year}
  {2006})}\BibitemShut {NoStop}%
\bibitem [{\citenamefont {Breuer}\ \emph {et~al.}(2016)\citenamefont {Breuer},
  \citenamefont {Laine}, \citenamefont {Piilo},\ and\ \citenamefont
  {Vacchini}}]{breuer2016colloquium}%
  \BibitemOpen
  \bibfield  {author} {\bibinfo {author} {\bibfnamefont {H.-P.}\ \bibnamefont
  {Breuer}}, \bibinfo {author} {\bibfnamefont {E.-M.}\ \bibnamefont {Laine}},
  \bibinfo {author} {\bibfnamefont {J.}~\bibnamefont {Piilo}},\ and\ \bibinfo
  {author} {\bibfnamefont {B.}~\bibnamefont {Vacchini}},\ }\href
  {https://doi.org/10.1103/RevModPhys.88.021002} {\bibfield  {journal}
  {\bibinfo  {journal} {Reviews of Modern Physics}\ }\textbf {\bibinfo {volume}
  {88}},\ \bibinfo {pages} {021002} (\bibinfo {year} {2016})}\BibitemShut
  {NoStop}%
\bibitem [{\citenamefont {Groeblacher}\ \emph {et~al.}(2015)\citenamefont
  {Groeblacher}, \citenamefont {Trubarov}, \citenamefont {Prigge},
  \citenamefont {Cole}, \citenamefont {Aspelmeyer},\ and\ \citenamefont
  {Eisert}}]{groeblacher2015observation}%
  \BibitemOpen
  \bibfield  {author} {\bibinfo {author} {\bibfnamefont {S.}~\bibnamefont
  {Groeblacher}}, \bibinfo {author} {\bibfnamefont {A.}~\bibnamefont
  {Trubarov}}, \bibinfo {author} {\bibfnamefont {N.}~\bibnamefont {Prigge}},
  \bibinfo {author} {\bibfnamefont {G.}~\bibnamefont {Cole}}, \bibinfo {author}
  {\bibfnamefont {M.}~\bibnamefont {Aspelmeyer}},\ and\ \bibinfo {author}
  {\bibfnamefont {J.}~\bibnamefont {Eisert}},\ }\href
  {https://doi.org/10.1038/ncomms8606} {\bibfield  {journal} {\bibinfo
  {journal} {Nature Communications}\ }\textbf {\bibinfo {volume} {6}},\
  \bibinfo {pages} {1} (\bibinfo {year} {2015})}\BibitemShut {NoStop}%
\bibitem [{\citenamefont {Triana}\ \emph {et~al.}(2016)\citenamefont {Triana},
  \citenamefont {Estrada},\ and\ \citenamefont
  {Pach{\'o}n}}]{triana2016ultrafast}%
  \BibitemOpen
  \bibfield  {author} {\bibinfo {author} {\bibfnamefont {J.~F.}\ \bibnamefont
  {Triana}}, \bibinfo {author} {\bibfnamefont {A.~F.}\ \bibnamefont
  {Estrada}},\ and\ \bibinfo {author} {\bibfnamefont {L.~A.}\ \bibnamefont
  {Pach{\'o}n}},\ }\href {https://doi.org/10.1103/PhysRevLett.116.183602}
  {\bibfield  {journal} {\bibinfo  {journal} {Physical Review Letters}\
  }\textbf {\bibinfo {volume} {116}},\ \bibinfo {pages} {183602} (\bibinfo
  {year} {2016})}\BibitemShut {NoStop}%
\bibitem [{\citenamefont {Zhang}\ \emph {et~al.}(2017)\citenamefont {Zhang},
  \citenamefont {Han}, \citenamefont {Xiong},\ and\ \citenamefont
  {Zhou}}]{zhang2017optomechanical}%
  \BibitemOpen
  \bibfield  {author} {\bibinfo {author} {\bibfnamefont {W.-Z.}\ \bibnamefont
  {Zhang}}, \bibinfo {author} {\bibfnamefont {Y.}~\bibnamefont {Han}}, \bibinfo
  {author} {\bibfnamefont {B.}~\bibnamefont {Xiong}},\ and\ \bibinfo {author}
  {\bibfnamefont {L.}~\bibnamefont {Zhou}},\ }\href
  {https://doi.org/10.1088/1367-2630/aa68d9} {\bibfield  {journal} {\bibinfo
  {journal} {New Journal of Physics}\ }\textbf {\bibinfo {volume} {19}},\
  \bibinfo {pages} {083022} (\bibinfo {year} {2017})}\BibitemShut {NoStop}%
\bibitem [{\citenamefont {Caldeira}\ and\ \citenamefont
  {Leggett}(1983{\natexlab{a}})}]{caldeira1983quantum}%
  \BibitemOpen
  \bibfield  {author} {\bibinfo {author} {\bibfnamefont {A.}~\bibnamefont
  {Caldeira}}\ and\ \bibinfo {author} {\bibfnamefont {A.}~\bibnamefont
  {Leggett}},\ }\href
  {https://doi.org/https://doi.org/10.1016/0003-4916(83)90202-6} {\bibfield
  {journal} {\bibinfo  {journal} {Annals of Physics}\ }\textbf {\bibinfo
  {volume} {149}},\ \bibinfo {pages} {374} (\bibinfo {year}
  {1983}{\natexlab{a}})}\BibitemShut {NoStop}%
\bibitem [{\citenamefont {Wei}\ and\ \citenamefont
  {Norman}(1963)}]{wei1963lie}%
  \BibitemOpen
  \bibfield  {author} {\bibinfo {author} {\bibfnamefont {J.}~\bibnamefont
  {Wei}}\ and\ \bibinfo {author} {\bibfnamefont {E.}~\bibnamefont {Norman}},\
  }\href {https://doi.org/10.1063/1.1703993} {\bibfield  {journal} {\bibinfo
  {journal} {Journal of Mathematical Physics}\ }\textbf {\bibinfo {volume}
  {4}},\ \bibinfo {pages} {575} (\bibinfo {year} {1963})}\BibitemShut {NoStop}%
\bibitem [{\citenamefont {Qvarfort}\ and\ \citenamefont
  {Pikovski}(2022)}]{qvarfort2022solving}%
  \BibitemOpen
  \bibfield  {author} {\bibinfo {author} {\bibfnamefont {S.}~\bibnamefont
  {Qvarfort}}\ and\ \bibinfo {author} {\bibfnamefont {I.}~\bibnamefont
  {Pikovski}},\ }\href {https://arxiv.org/abs/2210.11894} {\bibfield  {journal}
  {\bibinfo  {journal} {arXiv preprint arXiv:2210.11894}\ } (\bibinfo {year}
  {2022})}\BibitemShut {NoStop}%
\bibitem [{\citenamefont {Feynman}\ and\ \citenamefont
  {Vernon~Jr}(2000)}]{feynman2000theory}%
  \BibitemOpen
  \bibfield  {author} {\bibinfo {author} {\bibfnamefont {R.~P.}\ \bibnamefont
  {Feynman}}\ and\ \bibinfo {author} {\bibfnamefont {F.}~\bibnamefont
  {Vernon~Jr}},\ }\href {https://doi.org/10.1006/aphy.2000.6017} {\bibfield
  {journal} {\bibinfo  {journal} {Annals of physics}\ }\textbf {\bibinfo
  {volume} {281}},\ \bibinfo {pages} {547} (\bibinfo {year}
  {2000})}\BibitemShut {NoStop}%
\bibitem [{\citenamefont {Caldeira}\ and\ \citenamefont
  {Leggett}(1983{\natexlab{b}})}]{caldeira1983path}%
  \BibitemOpen
  \bibfield  {author} {\bibinfo {author} {\bibfnamefont {A.~O.}\ \bibnamefont
  {Caldeira}}\ and\ \bibinfo {author} {\bibfnamefont {A.~J.}\ \bibnamefont
  {Leggett}},\ }\href {https://doi.org/10.1016/0378-4371(83)90013-4} {\bibfield
   {journal} {\bibinfo  {journal} {Physica A: Statistical mechanics and its
  Applications}\ }\textbf {\bibinfo {volume} {121}},\ \bibinfo {pages} {587}
  (\bibinfo {year} {1983}{\natexlab{b}})}\BibitemShut {NoStop}%
\bibitem [{\citenamefont {Weiss}(2012)}]{weiss2012quantum}%
  \BibitemOpen
  \bibfield  {author} {\bibinfo {author} {\bibfnamefont {U.}~\bibnamefont
  {Weiss}},\ }\href@noop {} {\emph {\bibinfo {title} {Quantum dissipative
  systems}}},\ Vol.~\bibinfo {volume} {13}\ (\bibinfo  {publisher} {World
  scientific},\ \bibinfo {year} {2012})\BibitemShut {NoStop}%
\bibitem [{\citenamefont {Gardiner}\ and\ \citenamefont
  {Zoller}(2004)}]{gardiner2004quantum}%
  \BibitemOpen
  \bibfield  {author} {\bibinfo {author} {\bibfnamefont {C.}~\bibnamefont
  {Gardiner}}\ and\ \bibinfo {author} {\bibfnamefont {P.}~\bibnamefont
  {Zoller}},\ }\href@noop {} {\emph {\bibinfo {title} {Quantum noise: a
  handbook of Markovian and non-Markovian quantum stochastic methods with
  applications to quantum optics}}}\ (\bibinfo  {publisher} {Springer Science
  \& Business Media},\ \bibinfo {year} {2004})\BibitemShut {NoStop}%
\bibitem [{Note1()}]{Note1}%
  \BibitemOpen
  \bibinfo {note} {We see this by taking the commutator of the operators and
  finding that the results commute with all other operators.}\BibitemShut
  {Stop}%
\bibitem [{\citenamefont {Hu}\ \emph {et~al.}(2015)\citenamefont {Hu},
  \citenamefont {Huang}, \citenamefont {Liao}, \citenamefont {Tian},\ and\
  \citenamefont {Goan}}]{hu2015quantum}%
  \BibitemOpen
  \bibfield  {author} {\bibinfo {author} {\bibfnamefont {D.}~\bibnamefont
  {Hu}}, \bibinfo {author} {\bibfnamefont {S.-Y.}\ \bibnamefont {Huang}},
  \bibinfo {author} {\bibfnamefont {J.-Q.}\ \bibnamefont {Liao}}, \bibinfo
  {author} {\bibfnamefont {L.}~\bibnamefont {Tian}},\ and\ \bibinfo {author}
  {\bibfnamefont {H.-S.}\ \bibnamefont {Goan}},\ }\href
  {https://doi.org/10.1103/PhysRevA.91.013812} {\bibfield  {journal} {\bibinfo
  {journal} {Physical Review A}\ }\textbf {\bibinfo {volume} {91}},\ \bibinfo
  {pages} {013812} (\bibinfo {year} {2015})}\BibitemShut {NoStop}%
\bibitem [{\citenamefont {Qvarfort}\ \emph
  {et~al.}(2021{\natexlab{b}})\citenamefont {Qvarfort}, \citenamefont {Vanner},
  \citenamefont {Barker},\ and\ \citenamefont {Bruschi}}]{qvarfort2021master}%
  \BibitemOpen
  \bibfield  {author} {\bibinfo {author} {\bibfnamefont {S.}~\bibnamefont
  {Qvarfort}}, \bibinfo {author} {\bibfnamefont {M.~R.}\ \bibnamefont
  {Vanner}}, \bibinfo {author} {\bibfnamefont {P.~F.}\ \bibnamefont {Barker}},\
  and\ \bibinfo {author} {\bibfnamefont {D.~E.}\ \bibnamefont {Bruschi}},\
  }\href {https://doi.org/10.1103/PhysRevA.104.013501} {\bibfield  {journal}
  {\bibinfo  {journal} {Physical Review A}\ }\textbf {\bibinfo {volume}
  {104}},\ \bibinfo {pages} {013501} (\bibinfo {year}
  {2021}{\natexlab{b}})}\BibitemShut {NoStop}%
\bibitem [{\citenamefont {Zhang}\ \emph {et~al.}(2012)\citenamefont {Zhang},
  \citenamefont {Lo}, \citenamefont {Xiong}, \citenamefont {Tu},\ and\
  \citenamefont {Nori}}]{zhang2012general}%
  \BibitemOpen
  \bibfield  {author} {\bibinfo {author} {\bibfnamefont {W.-M.}\ \bibnamefont
  {Zhang}}, \bibinfo {author} {\bibfnamefont {P.-Y.}\ \bibnamefont {Lo}},
  \bibinfo {author} {\bibfnamefont {H.-N.}\ \bibnamefont {Xiong}}, \bibinfo
  {author} {\bibfnamefont {M.~W.-Y.}\ \bibnamefont {Tu}},\ and\ \bibinfo
  {author} {\bibfnamefont {F.}~\bibnamefont {Nori}},\ }\href
  {https://doi.org/10.1103/PhysRevLett.109.170402} {\bibfield  {journal}
  {\bibinfo  {journal} {Phys. Rev. Lett.}\ }\textbf {\bibinfo {volume} {109}},\
  \bibinfo {pages} {170402} (\bibinfo {year} {2012})}\BibitemShut {NoStop}%
\bibitem [{\citenamefont {Van~Rossum}\ and\ \citenamefont
  {Drake}(2009)}]{rossum2009python}%
  \BibitemOpen
  \bibfield  {author} {\bibinfo {author} {\bibfnamefont {G.}~\bibnamefont
  {Van~Rossum}}\ and\ \bibinfo {author} {\bibfnamefont {F.~L.}\ \bibnamefont
  {Drake}},\ }\href {https://docs.python.org/3/reference/} {\emph {\bibinfo
  {title} {Python 3 Reference Manual}}}\ (\bibinfo  {publisher} {CreateSpace},\
  \bibinfo {address} {Scotts Valley, CA},\ \bibinfo {year} {2009})\BibitemShut
  {NoStop}%
\bibitem [{\citenamefont {Virtanen}\ \emph {et~al.}(2020)\citenamefont
  {Virtanen}, \citenamefont {Gommers}, \citenamefont {Oliphant}, \citenamefont
  {Haberland}, \citenamefont {Reddy}, \citenamefont {Cournapeau}, \citenamefont
  {Burovski}, \citenamefont {Peterson}, \citenamefont {Weckesser},
  \citenamefont {Bright}, \citenamefont {{van der Walt}}, \citenamefont
  {Brett}, \citenamefont {Wilson}, \citenamefont {Millman}, \citenamefont
  {Mayorov}, \citenamefont {Nelson}, \citenamefont {Jones}, \citenamefont
  {Kern}, \citenamefont {Larson}, \citenamefont {Carey}, \citenamefont {Polat},
  \citenamefont {Feng}, \citenamefont {Moore}, \citenamefont {{VanderPlas}},
  \citenamefont {Laxalde}, \citenamefont {Perktold}, \citenamefont {Cimrman},
  \citenamefont {Henriksen}, \citenamefont {Quintero}, \citenamefont {Harris},
  \citenamefont {Archibald}, \citenamefont {Ribeiro}, \citenamefont
  {Pedregosa}, \citenamefont {{van Mulbregt}},\ and\ \citenamefont {{SciPy 1.0
  Contributors}}}]{virtanen2020scipy}%
  \BibitemOpen
  \bibfield  {author} {\bibinfo {author} {\bibfnamefont {P.}~\bibnamefont
  {Virtanen}}, \bibinfo {author} {\bibfnamefont {R.}~\bibnamefont {Gommers}},
  \bibinfo {author} {\bibfnamefont {T.~E.}\ \bibnamefont {Oliphant}}, \bibinfo
  {author} {\bibfnamefont {M.}~\bibnamefont {Haberland}}, \bibinfo {author}
  {\bibfnamefont {T.}~\bibnamefont {Reddy}}, \bibinfo {author} {\bibfnamefont
  {D.}~\bibnamefont {Cournapeau}}, \bibinfo {author} {\bibfnamefont
  {E.}~\bibnamefont {Burovski}}, \bibinfo {author} {\bibfnamefont
  {P.}~\bibnamefont {Peterson}}, \bibinfo {author} {\bibfnamefont
  {W.}~\bibnamefont {Weckesser}}, \bibinfo {author} {\bibfnamefont
  {J.}~\bibnamefont {Bright}}, \bibinfo {author} {\bibfnamefont {S.~J.}\
  \bibnamefont {{van der Walt}}}, \bibinfo {author} {\bibfnamefont
  {M.}~\bibnamefont {Brett}}, \bibinfo {author} {\bibfnamefont
  {J.}~\bibnamefont {Wilson}}, \bibinfo {author} {\bibfnamefont {K.~J.}\
  \bibnamefont {Millman}}, \bibinfo {author} {\bibfnamefont {N.}~\bibnamefont
  {Mayorov}}, \bibinfo {author} {\bibfnamefont {A.~R.~J.}\ \bibnamefont
  {Nelson}}, \bibinfo {author} {\bibfnamefont {E.}~\bibnamefont {Jones}},
  \bibinfo {author} {\bibfnamefont {R.}~\bibnamefont {Kern}}, \bibinfo {author}
  {\bibfnamefont {E.}~\bibnamefont {Larson}}, \bibinfo {author} {\bibfnamefont
  {C.~J.}\ \bibnamefont {Carey}}, \bibinfo {author} {\bibfnamefont
  {{\.I}.}~\bibnamefont {Polat}}, \bibinfo {author} {\bibfnamefont
  {Y.}~\bibnamefont {Feng}}, \bibinfo {author} {\bibfnamefont {E.~W.}\
  \bibnamefont {Moore}}, \bibinfo {author} {\bibfnamefont {J.}~\bibnamefont
  {{VanderPlas}}}, \bibinfo {author} {\bibfnamefont {D.}~\bibnamefont
  {Laxalde}}, \bibinfo {author} {\bibfnamefont {J.}~\bibnamefont {Perktold}},
  \bibinfo {author} {\bibfnamefont {R.}~\bibnamefont {Cimrman}}, \bibinfo
  {author} {\bibfnamefont {I.}~\bibnamefont {Henriksen}}, \bibinfo {author}
  {\bibfnamefont {E.~A.}\ \bibnamefont {Quintero}}, \bibinfo {author}
  {\bibfnamefont {C.~R.}\ \bibnamefont {Harris}}, \bibinfo {author}
  {\bibfnamefont {A.~M.}\ \bibnamefont {Archibald}}, \bibinfo {author}
  {\bibfnamefont {A.~H.}\ \bibnamefont {Ribeiro}}, \bibinfo {author}
  {\bibfnamefont {F.}~\bibnamefont {Pedregosa}}, \bibinfo {author}
  {\bibfnamefont {P.}~\bibnamefont {{van Mulbregt}}},\ and\ \bibinfo {author}
  {\bibnamefont {{SciPy 1.0 Contributors}}},\ }\href
  {https://doi.org/10.1038/s41592-019-0686-2} {\bibfield  {journal} {\bibinfo
  {journal} {Nature Methods}\ }\textbf {\bibinfo {volume} {17}},\ \bibinfo
  {pages} {261} (\bibinfo {year} {2020})}\BibitemShut {NoStop}%
\bibitem [{\citenamefont {\textit{et al}.}(2010)}]{mpmath}%
  \BibitemOpen
  \bibfield  {author} {\bibinfo {author} {\bibfnamefont {F.~J.}\ \bibnamefont
  {\textit{et al}.}},\ }\href {http://code.google.com/p/mpmath/} {\emph
  {\bibinfo {title} {mpmath: a {P}ython library for arbitrary-precision
  floating-point arithmetic (version 0.14)}}} (\bibinfo {year}
  {2010})\BibitemShut {NoStop}%
\end{thebibliography}%

\end{document}